\documentclass[a4paper,11pt]{article}
\pdfoutput=1 

\usepackage{jcappub} 
\usepackage{lineno}
\usepackage[normalem]{ulem}
\usepackage[utf8]{inputenc}
\usepackage{enumerate}
\usepackage{amsmath, amssymb, amsthm, graphicx, epsfig, fancyhdr,epsfig, slashed}
\usepackage[mathscr]{euscript}
\let\oldcdot\cdot
\usepackage{breqn}
\let\cdot\oldcdot
\usepackage{epsfig}  
\usepackage{graphicx}   
\usepackage{slashed}       
\usepackage{tikz}
\usepackage{subcaption} 
\usepackage{url}
\usepackage{color,soul} 
\usepackage{xcolor}
\usepackage{multirow} 
\usepackage{comment}
\usepackage{mathtools}
\usepackage{bm}
\usepackage{hyperref}
\usepackage[all]{hypcap}
\hypersetup{  
    colorlinks=true,
    linkcolor=blue,
    filecolor=blue,      
    urlcolor=blue,
    citecolor=blue,
    }

\title{Tracing Ultra Light Axions in Post-reionization, Lyman-$\alpha$ and CMB Missions}

\author[a]{Debarun Paul,}
\author[a]{Sourav Pal,}
\author[a]{Amit Dutta Banik,}
\author[a]{Supratik Pal}

\affiliation[a]{\,Physics and Applied Mathematics Unit, Indian Statistical Institute, 203 B.T. Road, Kolkata 700108, India}

\emailAdd{debarun31paul@gmail.com}
\emailAdd{soupal1729@gmail.com}
\emailAdd{amitdbanik@gmail.com}
\emailAdd{supratik@isical.ac.in}

\abstract{Ultra-light axions (ULAs) are  dark matter candidates proposed to resolve the small scale anomalies of the standard cosmological model. Due to their inherent quantum pressure, ULAs result in a distinct, scale-dependent suppression on the matter power spectrum, which can leave imprints on the upcoming observations. We explore such possibilities by forecasting on the post-reionization large scale structure (LSS) surveys and next-generation cosmic microwave background (CMB) missions. By utilizing the cross-correlation between 21-cm intensity mapping (SKA1-MID and PUMA) and the Lyman-$\alpha$ forest (DESI-like), we explore possible signatures of ULAs in post-reionization surveys while  mitigating instrument-specific systematics.  The Fisher matrix analysis projects uncertainties on the fractional ULA abundance across a wide ULA mass range of $10^{-30}\text{ eV} \le m_a \le 10^{-20}\text{ eV}$, revealing an optimal detection sensitivity at intermediate masses around $m_a \sim 10^{-25}\text{ eV}$. Furthermore, while  next-generation  CMB mission alone can yield small projected errors on the ULA fraction compared to future LSS missions, a joint analysis of the DESI-like and PUMA cross-spectrum alongside CMB-S4-like missions estimates an error on the ULA fraction to be $\mathcal{O}(10^{-4})$ for $m_a\lesssim 10^{-28}$ eV, highlighting a significant improvement over standalone LSS and CMB missions.}
\begin{document}
\maketitle
\flushbottom

\section{Introduction}
\label{sec:introduction}

The nature of dark matter (DM) remains one of the mysteries in modern cosmology. While the standard $\Lambda\mathrm{CDM}$ (Cold Dark Matter with a cosmological constant, $\Lambda$) model successfully explains the large scale structure (LSS) of the Universe~\cite{Bahcall:1999xn}, it faces significant challenges at the small scales~\citep{Flores:1994gz,Moore:1994yx,Moore:1999gc,Moore:1999nt,Klypin:1999uc,Boylan-Kolchin:2011qkt,Boylan-Kolchin:2011lmk}. To resolve these discrepancies, an alternative paradigm suggests that dark matter may consist of light bosons, such as ultra-light axions (ULAs) or axion-like particles (ALPs)~\cite{Bauer:2020zsj, Hlozek:2017zzf}. These models address the aforementioned issues through the formation of solitonic cores and the suppression of small scale structure. Originating as pseudo-Nambu-Goldstone bosons, ULAs span an exceptionally light mass range ($m_a \sim 10^{-30}\text{--}10^{-20}~\mathrm{eV}$)~\cite{Hlozek:2014lca, Marsh:2015xka}. Despite their extremely small masses, ULAs can behave like standard CDM on large scales~\cite{Hwang:2009js} due to their non-thermal production in the early Universe via the misalignment mechanism~\cite{Preskill:1982cy, Dine:1982ah, Abbott:1982af,Sikivie:2006ni,Marsh:2015xka}. However, their de Broglie wavelengths generate a quantum pressure that restricts gravitational clustering, inherently suppressing small scale structure. Furthermore, recent studies highlight their capability to relax both the $H_0$~\cite{Blum:2021oxj} and $\sigma_8$~\cite{Allali:2021azp,Lague:2021frh,Ye:2021iwa} tensions. This small scale suppression leaves distinct scale-dependent imprints on the matter power spectra, offering a unique opportunity to probe the microphysical nature of dark matter~\cite{Hlozek:2017zzf, Lague:2021frh}.

In this context, the post-reionization era ($z \lesssim 6$) emerges as an ideal laboratory to probe this mildly non-linear small scale suppression. The small scale cutoff characteristic of ULA physics lies precisely in the regime accessible to complementary high-redshift intergalactic medium (IGM) tracers, most notably the 21-cm intensity mapping signal and the Lyman-$\alpha$ ($\mathrm{Ly}\alpha$) forest~\cite{Bauer:2020zsj, Lague:2021frh}. The 21-cm hyperfine transition of neutral hydrogen ($\mathrm{HI}$) provides a wide-field 3D map of the large scale matter distribution~\cite{Bull:2014rha}. By measuring the integrated emission from large cosmological volumes without resolving individual galaxies, upcoming radio interferometers like the Square Kilometre Array (specifically, SKA1-MID)~\footnote{\url{https://www.skao.int/en}}~\cite{SKA:2018ckk} and the proposed Packed Ultrawideband Mapping Array (PUMA)~\footnote{\url{https://www.puma.bnl.gov/}}~\cite{CosmicVisions21cm:2018rfq} can efficiently map the matter power spectrum across a wide range of redshifts and scales. Complementing these wide-field radio observations, the $\mathrm{Ly}\alpha$ forest captures small scale density fluctuations ($k \sim 0.1\text{--}10~h\,\mathrm{Mpc}^{-1}$) via resonant photon absorption along quasar sight-lines~\cite{McDonald:2006qs}, which will be observed by next-generation optical spectroscopic surveys like the Dark Energy Spectroscopic Instrument (DESI)~\footnote{\url{https://www.desi.lbl.gov/}}~\cite{DESI:2024mwx}. Although the analyses by Dark Energy Survey (DES)~\footnote{\url{https://www.darkenergysurvey.org/}} Year 1 data have put questions on ULAs as the sole component of dark matter~\cite{Dentler:2021zij}, ULAs can still constitute a significant fractional component of the total DM cosmic inventory, thereby respecting the constraints from DES Y1. For $10^{-23}~\mathrm{eV}\lesssim m_a\lesssim 10^{-21}$~eV, the fractional abundance of ULAs (\textit{i.e.} $\Omega_a/\Omega_{\rm dm}$, where $\Omega_a$ presents the relic abundance of ULAs and $\Omega_{\rm dm}$ is the relic abundance of the total matter) is constrained to an upper bound of approximately $0.1$ by $\mathrm{Ly}\alpha$ observations~\cite{Irsic:2017yje,Armengaud:2017nkf,Rogers:2020ltq,Kobayashi:2017jcf}. Recent analysis puts a tighter upper limit of $\lesssim 0.007$ for $m_a=10^{-23}$ eV~\cite{Liu:2026jkq}. For lighter masses, $10^{-26}~\mathrm{eV} \lesssim m_a \lesssim 10^{-23}$~eV, the projected upper limit becomes significantly tighter, reaching $\sim 0.01$ based on forecasts for future 21-cm measurements by the Hydrogen Epoch of Reionization Array (HERA)~\footnote{\url{https://reionization.org/}}~\cite{Lazare:2024uvj}.

Despite the immense promise shown by these individual measurements, their full observational potential requires a methodical, multi-tracer framework. In addition to utilizing the individual 21-cm and $\mathrm{Ly}\alpha$ auto-correlations, the cross-correlation between these two probes serves as a remarkably clean and powerful cosmological diagnostic~\cite{Carucci:2016yzq,CHIME:2023til,Montero-Camacho:2024xvf}. Because 21-cm intensity mapping relies on radio interferometry and the $\mathrm{Ly}\alpha$ forest on optical spectroscopy, their respective instrumental systematics and primary foreground contaminants are expected to be entirely statistically independent. Consequently, the cross-power spectrum cleanly isolates the shared underlying DM distribution while heavily suppressing these individual systematic effects, a feature that has been highlighted in recent studies probing background cosmological models~\cite{Bhaumik:2026mlc,Pal:2026cgj}. In the context of ULA DM, this cross-correlation retains the distinctive, scale-dependent morphological imprint of the small scale power suppression driven by quantum pressure. By methodically exploring this observational potential, the present study aims to establish a joint constraints that fully harness the complementary strengths of the post-reionization era. 

Apart from LSS observations, the cosmic microwave background (CMB) plays a crucial role in probing ULA DM. Because ULAs alter the effective matter content of the Universe around the epoch of matter-radiation equality, they strongly affect the amplitude of the acoustic peaks. For ULAs of mass  $10^{-32}~\mathrm{eV}\lesssim m_a \lesssim 10^{-29}$ eV and fractional ULA abundance $\gtrsim 0.5$, matter-radiation equality may get significantly delayed, which may affect the amplitude of the Integrated Sachs-Wolfe (ISW)~\cite{Hlozek:2014lca, Coble:1996te}. However,  Planck-18 data~\cite{Planck:2018vyg}, in conjunction with the Baryon Oscillation Spectroscopic Survey (BOSS)~\cite{eBOSS:2020yzd}, has constrained the fractional abundance of ULAs in the mass range $10^{-28}~\mathrm{eV} \lesssim m_a \lesssim 10^{-26}$~eV to $\lesssim 0.01$~\cite{Lague:2021frh}, thereby reducing its prospects in the ISW effect to a moderate level. Furthermore, a recent analysis incorporating Planck-18+DESI DR2 provides an even tighter upper limit of $\lesssim 0.007$ on fractional ULA abundance, within $10^{-30}~\mathrm{eV} \lesssim m_a \lesssim 10^{-28}$ eV~\cite{Liu:2025qwf}. Forecasts for the high-sensitivity Stage 4 CMB missions (for example, CMB-S4) anticipate tightening this upper limit further to $\sim 0.001$~\cite{Hlozek:2016lzm}.

Motivated by these prospects, the primary goal of this study is to provide projected limits on the fractional abundance of axions using the synergy of post-reionization 21-cm and $\mathrm{Ly}\alpha$ cross-correlations in combination with high-sensitivity CMB-S4-like future experiments. We model the axion fluid perturbations in the mildly non-linear regime using the publicly available Boltzmann solver \texttt{AxiCLASS}~\footnote{Publicly available at: \url{https://github.com/PoulinV/AxiCLASS}}~\cite{Smith:2019ihp, Poulin:2018dzj} to track the suppression signatures accurately. We then perform a comprehensive Fisher matrix forecasting analysis for a DESI-like spectroscopic survey, the SKA1-MID single-dish configuration and the PUMA interferometer, alongside CMB-S4-like future missions~\cite{CMB-S4:2016ple}. A central result of our analysis demonstrates that the inclusion of multi-tracer cross-correlations effectively breaks standard cosmological parameter degeneracies, significantly tightening the marginalized exclusion limits on the ULA parameter space relative to a CMB-only baseline.

The paper is organised as follows. In Sec.~\ref{sec:DMmodel}, we briefly review the microphysical framework of ULA DM and its background and perturbative cosmological implications. In Sec.~\ref{sec:power_spectra}, we outline the analytical framework for the post-reionization 21-cm power spectrum, the $\mathrm{Ly}\alpha$ flux power spectrum and their cross-correlation of the underlying matter distribution. In Sec.~\ref{sec:detection_prospect}, we present the estimation of the signal-to-noise ratio (SNR) and the joint parameter constraints obtained from our Fisher matrix analysis. Finally, we summarize our key findings and discuss future directions for large scale structure multi-tracer observations in Sec.~\ref{sec:discussion}.

\section{Ultra-light axion dark matter}
\label{sec:DMmodel}
In this section, we briefly review the ULAs and discuss their cosmological implications. ULAs are pseudo-Numbo-Goldstone bosons, typically modelled as scalar fields ($\phi$). Their dynamics are governed by a cosine potential~\cite{Marsh:2015xka,Yang:2017wzh}
\begin{eqnarray}
    \label{eq:axion_pot}
    V(\phi) = \Lambda^4\left[1-\cos\left(\frac{\phi}{f_a}\right)\right],
\end{eqnarray}
where $f_a$ is the $U(1)$ symmetry breaking scale. The energy scale $\Lambda$ is associated with the mass of axions ($m_a$) via $\Lambda^4\approx m_a^2 f_a^2$. The background evolution of the scalar field in a Friedmann-Lema\^{i}tre-Robertson-Walker (FLRW) Universe is governed by the Klein-Gordon equation
\begin{equation}
    \phi^{\prime\prime} + 2\mathcal{H}\phi^{\prime} + \frac{dV(\phi)}{d\phi} = 0 \,,
\end{equation}
where primes denote derivatives with respect to conformal time $\eta$ and $\mathcal{H} \equiv a^{\prime}/a$ is the conformal Hubble expansion rate, with $a(\eta)$ being the scale factor. In general, it is assumed that $\phi \ll f_a$~\cite{Bauer:2020zsj,Hlozek:2017zzf,Lague:2021frh}~\footnote{There are specific scenarios where $\phi\sim f_a$ can have distinctive cosmological implications as explored in~\cite{Hlozek:2014lca}.}, allowing us to approximate the axion potential (Eq.~\eqref{eq:axion_pot}) as $V(\phi)\approx \frac{1}{2}m_a^2\phi^2$.
When the Hubble parameter drops to $H\sim m_a$, the axion field starts to oscillate, corresponding to the scale factor $a_{\rm osc}$. At the latter time ($H\ll m_a$), the field oscillates rapidly with a frequency $\sim m_a$, and its time-averaged energy density redshifts as matter, \textit{i.e.} $\langle \rho_a \rangle\propto a^{-3}$, effectively mimicking standard CDM on large scales.

For ULAs, the de Broglie wavelength is very large due to its exceptionally low mass. Thus, the DM fluid behaves as a coherent scalar field governed by the Schr\"odinger-Poisson equations~\cite{Chadha-Day:2021szb,OHare:2024nmr,Poulin:2018dzj}. The resistance to wavepacket localization, dictated by the Heisenberg uncertainty principle, manifests as a strictly scale-dependent effective sound speed~\cite{Hwang:2009js}
\begin{eqnarray}
\label{eq:cs2}
    c_s = \left[\frac{\frac{k^2}{4m_a^2 a^2}}{1 + \frac{k^2}{4m_a^2 a^2}}\right]^{1/2} \,,
\end{eqnarray}
with $k$ being the wavenumber. This quantum pressure introduces a characteristic `Jeans' scale, $k_{\rm J}$, above which (\textit{i.e.} $k>k_{\rm J}$) gravitational clustering is suppressed~\footnote{While the scale is observationally relevant for ULAs, for QCD (quantum chromodynamics) axions~\cite{Preskill:1982cy,Abbott:1982af,Kim:2008hd,Peccei:2006as}, the much larger mass ($m_a\sim 10^{-6}$ eV) render the cutoff scale cosmologically unobservable.}. The Jeans scale can be estimated as $c_s^2 k_{\rm J}^2=4\pi Ga^2\rho_a^2$~\cite{Khlopov:1985fch}, where $G$ is the Newton's gravitational constant and $\rho_a$ is the energy density of the axions. Consequently, on large scales ($k<k_{\rm J}$), the quantum pressure is negligible, and the fluid behaves like CDM. Conversely, on the small scales ($k>k_{\rm J}$) quantum pressure prevents gravitational clustering, leading to a suppression of the matter power spectrum. At the late time ($a\gg a_{\rm osc}$), the density and velocity perturbation equations for the axion fluid in synchronous gauge can be expressed as~\cite{Hlozek:2014lca,Marsh:2015xka}
\begin{eqnarray}
    \delta_a^{\prime} &=& -k u_a - \frac{h^{\prime}}{2} - 3\mathcal{H}c_s^2\delta_a - 9\mathcal{H}^2 c_s^2 u_a/k \,, \label{eq:axion_pert1} \\
    u_a^{\prime} &=& -\mathcal{H}u_a + c_s^2 k \delta_a + 3\mathcal{H}c_s^2 u_a \,\label{eq:axion_pert2},
\end{eqnarray}
where $u_a$ is the dimensionless perturbed heat flux. The additional  terms on the right hand side of the equations are solely due to the non-zero effective sound speed. For relativistic modes ($k>m_a a$), the effective sound speed is dominant as it approaches $c_s \to 1$ (\textit{e.g.} Eq.~\eqref{eq:cs2}). However, for non-relativistic modes $k<m_a a$, the effective sound speed (Eq.~\eqref{eq:cs2}) reduces to $c_s\approx \frac{k}{2m_a a}$. For the very small modes ($k\ll m_a a$) \textit{i.e.} the large scales, the effective sound speed vanishes ($c_s\to 0$), recovering standard CDM dynamics, as evident from Eqs.~\eqref{eq:axion_pert1}-\eqref{eq:axion_pert2}. Throughout our analysis, we used publicly available Boltzmann solver for axion \texttt{AxiCLASS}~\cite{Smith:2019ihp,Poulin:2018dzj,Poulin:2018cxd,Murgia:2020ryi}, to perform all numerical calculations.

\subsection{Cosmological implications of the model}
\label{subsec:DM_cosmology}
Having established the underlying phenomenology, we now demonstrate the signature of ULAs on matter power spectrum and CMB. The cosmological impact of ULAs is primarily governed by their mass ($m_a$) and fractional abundance ($\Omega_a/\Omega_{\rm dm}$), where $\Omega_a$ presents the relic abundance of ULAs and $\Omega_{\rm dm}=0.26067$ is the relic abundance of the total matter content of the Universe which is fixed at its best fit value of Planck TT, TE, EE+lowE+lensing+BAO 2018~\cite{Planck:2018vyg}. To highlight the impacts of the ULA-parameters, we set the standard six cosmological parameters~\footnote{By standard six cosmological parameters, we refer: present-day relic abundance of baryon ($\omega_{b}\equiv \Omega_b h^2$) and CDM ($\omega_{\rm cdm}\equiv \Omega_{\rm cdm} h^2$), reduced Hubble parameter ($h$), amplitude of scalar perturbations ($A_s$) at k=0.05 Mpc$^{-1}$, scalar spectral index ($n_s$) and optical depth during the epoch of reionization ($\tau$).} to the best fit values of the Planck-18 data~\cite{Planck:2018vyg}. In the subsequent power spectra, we illustrate the impact of varying the ULA parameters individually while holding the other constant. Furthermore, where ULAs are contributing as a fraction of DM, the total DM abundance ($\Omega_{\rm dm}$) is conserved by the CDM component ($\Omega_{\rm cdm}$), \textit{i.e.} $\Omega_{\rm dm} = \Omega_a + \Omega_{\rm cdm}$.   
\begin{figure}[!ht]
    \centering
    \includegraphics[width=1.0\linewidth]{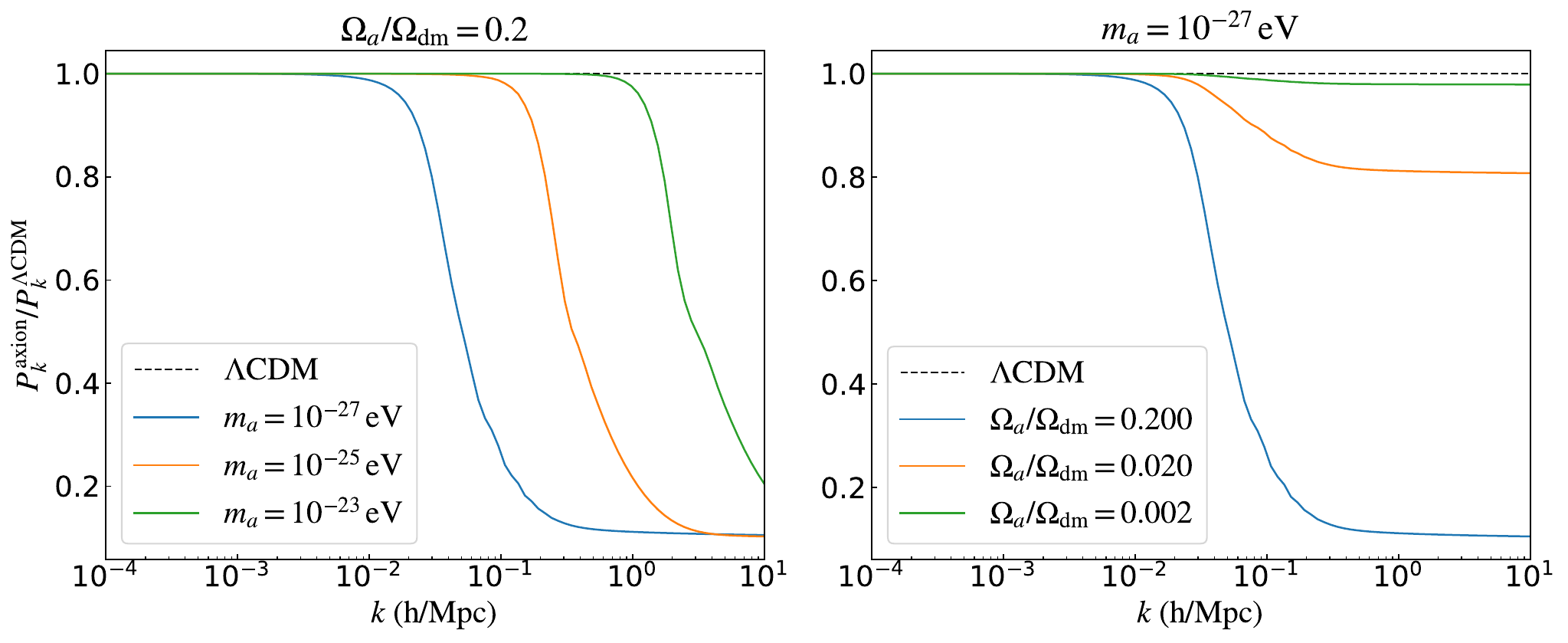}
    \caption{\it Illustration of the ratio of ULA matter power spectrum to the standard $\Lambda$CDM prediction. Each panel showcases the impact of a single ULA parameter while keeping the other fixed. \textbf{Left panel} presents the variation of the axion mass ($m_a$) for $\Omega_a/\Omega_{\rm dm}=0.2$, showing suppressions at the small scale. The suppressions shift to the smaller $k$ for lighter masses. \textbf{Right panel} presents the variation of the axion fractional abundance ($\Omega_a/\Omega_{\rm dm}$) for $m_a=10^{-27}$ eV. A larger axion fraction have a more significant small scale suppression, beginning the suppression at the same $k$.}
    \label{fig:axion_PS}
\end{figure}

In Fig.~\ref{fig:axion_PS}, we present the impact of $m_a$ (left panel) and $\Omega_a$ (right panel) on matter power spectra, normalised by $\Lambda$CDM prediction (black dashed lines). In the left panel,  the fraction of axion abundance is fixed to $\Omega_a/\Omega_{\rm dm}=0.2$ and vary $m_a$. A significant suppression at the small scales (large $k$) due to the aforementioned quantum pressure can be observed from the plots.   The suppression sets in at $k=k_{\rm J}$, which increases with the increase of $m_a$. Consequently, for lighter axions, the suppression begins relatively at smaller $k$, \textit{e.g.} for $m_a=10^{-27}$ eV, $k_{\rm J}\approx 10^{-2}$ h/Mpc, whereas $k_{\rm J}\approx 1$ h/Mpc for $m_a=10^{-23}$ eV. The right panel demonstrates the impacts of fraction of axion abundance at $m_a=10^{-27}$ eV. The suppression becomes more prominent as the fraction of axion abundance increases, recovering the pure CDM limit as $\Omega_a/\Omega_{\rm dm}\to 0$. As the Jeans scale is predominantly dictated by the mass of axions, the suppression begins relatively at same $k$ across all $\Omega_a/\Omega_{\rm dm}$, which is approximately $10^{-2}$ h/Mpc as $m_a=10^{-27}$ eV. This highlights the non-degeneracy between these two parameters. 
\begin{figure}[!ht]
    \centering
    \includegraphics[width=1.0\linewidth]{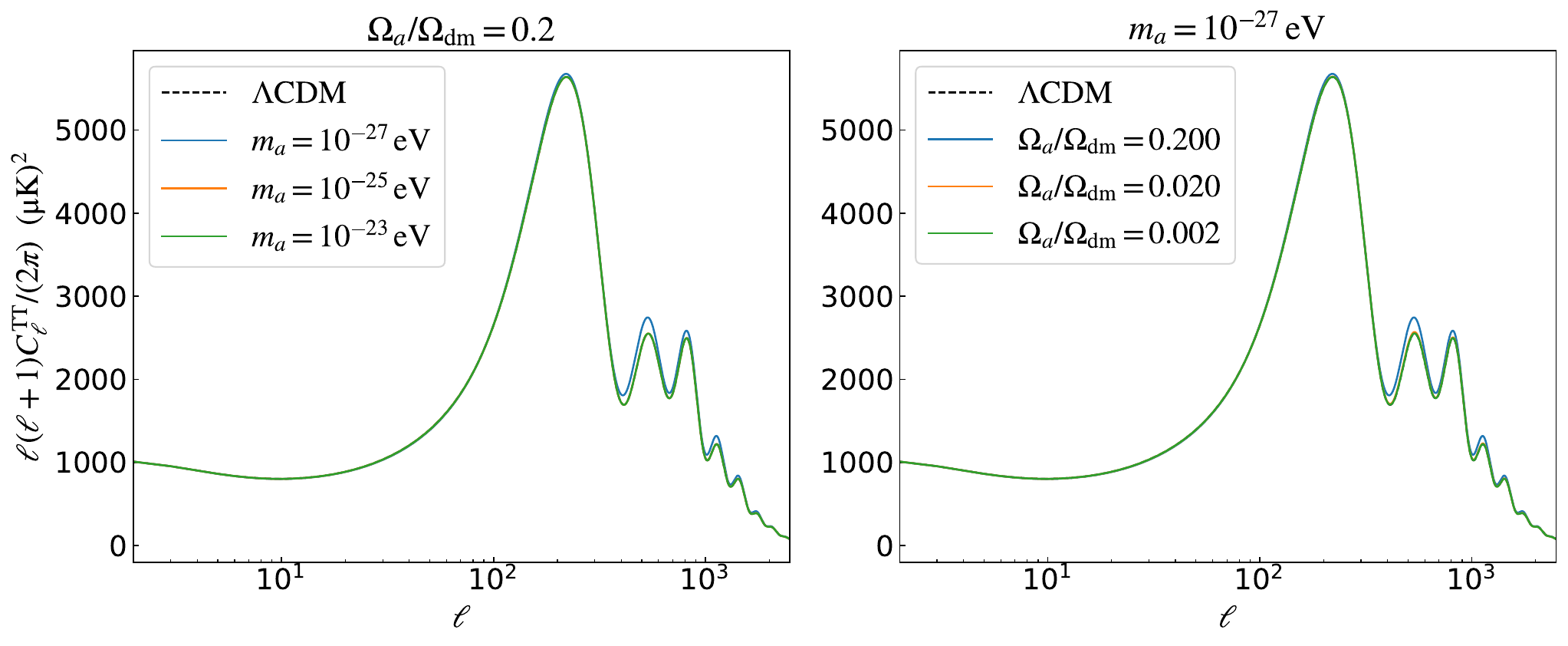}
    \caption{\it Illustration of CMB TT power spectrum for ULAs. Black dashed line presents the $\Lambda$CDM prediction. Each panel showcases the impact of a single ULA parameter while keeping the other fixed. \textbf{Left panel} presents the variation of $m_a$ for $\Omega_a/\Omega_{\rm dm}=0.2$, whereas \textbf{right panel} shows the variation of $\Omega_a/\Omega_{\rm dm}$ for fixed $m_a=10^{-27}$ eV. A smaller axion mass or larger axion fraction systematically alters the acoustic peaks. In both scenarios, the deviations are driven by non-zero sound speed of the ULAs, while the horizontal positions of the peaks remain largely unchanged.}
    \label{fig:axion_CMB}
\end{figure}

Let us now turn to the CMB power spectrum where we investigate the impacts of ULA-parameters on the two-point temperature auto correlation (TT) of the CMB power spectra. As clustering is suppressed at the small scales for the ULAs, gravitational drag on the photon-baryon plasma is largely affected, leaving distinct imprints on the acoustic peaks.

Fig.~\ref{fig:axion_CMB} depicts how the mass and fraction of axion DM alter the CMB TT power spectrum compared to $\Lambda$CDM scenario (black dashed line). Since the ULAs, having mass $m_a\gtrsim 10^{-28}$ eV, behave as matter during the matter-domination era, the low $\ell$ of CMB TT power spectrum (Sachs-Wolfe plateau) remains unaltered compared to CDM scenario.  Additionally, since the angular size of the sound horizon remains largely unaltered, the positions of the acoustic peaks do not shift significantly compared to CDM scenarios. However, as the gravitational drag on the photon-baryon plasma is affected, the amplitudes of the peaks get altered. The left panel demonstrates the variation of $m_a$ with a fixed fraction $\Omega_a/\Omega_{\rm dm}=0.2$, while the right panel presents the variation of $\Omega_a/\Omega_{\rm dm}$, keeping $m_a=10^{-27}$ eV. The left panel showcases that for $m_a \lesssim 10^{-25}$ eV CMB power spectra for ULAs mostly indistinguishable from $\Lambda$CDM prediction. However, for $m_a=10^{-27}$ eV, the onset of axion oscillations coincides roughly with the epoch of matter-radiation equality, causing a slight delay in its timing. This results into an enhancement of the first peak due to the lack of gravitational pulling in the photon-baryon plasma. Furthermore, the lack of gravitational pulling leads to the suppression of the third peak (corresponding to the second compression phase of the photon-baryon plasma) compared to the second acoustic peak. The right panel of the figure corroborates the same physical significance. Larger the axion abundance enhances the first acoustic peak, along with the suppression of the third peak compared to the second peak. As expected, the limit $\Omega_a/\Omega_{\rm dm}\to 0$ recovers the standard $\Lambda$CDM scenario. 

Such distinct cosmological imprints of ULAs on both the matter and CMB power spectra present a compelling case to investigate the scenario further in a concrete way, leading to possible bounds on the corresponding parameter space. 
In this context, the intermediate tracers of the IGM may serve as the window into new physics. 
Specifically, the Ly$\alpha$ forest observed by facilities like the Dark Energy Spectroscopic Instrument (DESI) and the 21-cm signal from the post-reionization targeted by the Square Kilometre Array (SKA)~\cite{2015aska.confE..86N,SKA:2018ckk,Weltman:2018zrl} and Packed Ultrawideband Mapping Array (PUMA)~\cite{CosmicVisions21cm:2018rfq,PUMA:2019jwd,Castorina:2020zhz} offer unprecedented sensitivity to the small scale power spectrum. Simultaneously, altering the amplitudes of acoustic peaks of CMB TT spectrum demands the temperature and polarization data anticipated from the upcoming Stage 4 CMB missions. Motivated by these observational prospects, the subsequent sections systematically investigates the ULA signatures on the 21-cm and Ly$\alpha$ power spectra, as well as their cross power spectra, culminating with a robust forecasting analysis to project the joint constraining power of these future LSS missions in synergy with CMB-S4-like future CMB experiments.

\section{Signature of the dark matter on 21-cm and Lyman-\texorpdfstring{$\alpha$}{} power spectrum}
\label{sec:power_spectra}
In this section, we detail the analytical framework for the Ly$\alpha$ forest and 21-cm power spectra employed in our forecast analysis. As discussed, ULAs can have non-trivial effects on the early universe modifies the growth of structures, resulting in a characteristic, scale-dependent suppression of the matter power spectrum. During the post-reionization era ($z \lesssim 6$), these axionic imprints are expected to be more pronounces and can be effectively mapped using the Ly$\alpha$ forest and 21-cm intensity mapping signals. Both observables are modeled as biased tracers of the underlying DM distribution, incorporating the necessary transformations into redshift space to account for peculiar velocities.
\subsection{21-cm power spectra}
\label{subsec:21cm-ps}
In the post-reionization epoch, the three-dimensional (3D) power spectrum of 21-cm fluctuations serves as an effective tracer for the aggregate emission of neutral hydrogen (HI). By incorporating redshift space distortions (RSD), specifically the linear Kaiser effect and the non-linear Fingers-of-God (FoG) damping, the observed 21-cm power spectrum is modelled as~\cite{Bull:2014rha,Karagiannis:2020dpq}
\begin{equation}
\label{eq:21_ps}
    P_{21}^{\mathrm{3D}}(\boldsymbol{k},z) = \bar{T}_{21}(z)^2 b_{\mathrm{HI}}^2(k,z) \left(1+\beta_{\mathrm{HI}}\mu^2\right)^2 P_{\mathrm{m}}(k,z) D_{\mathrm{FoG}}(k_{\parallel}, \sigma_{v,21})\, ,
\end{equation}
where $P_{\mathrm{m}}(k,z)$ represents the underlying matter power spectrum. The global 21-cm brightness temperature, $\bar{T}_{21}(z)$, is determined by the cosmological HI density parameter $\Omega_{\rm HI}(z)$:
\begin{equation}\label{eq:Tbar}
    \bar{T}_{21}(z) = 188\,\frac{h}{E(z)}\,\Omega_{\rm HI}(z)\,(1+z)^2\;\rm mK \, ,
\end{equation}
with $E(z) = H(z)/H_0$ being the dimensionless expansion rate. We adopt the evolution $\Omega_{\rm HI}(z) = 4\times 10^{-4}(1+z)^{0.6}$ to characterize the HI content across the redshifts of interest.

The scale-dependent bias parameter, $b_{\mathrm{HI}}(k,z)$, controls the spatial distribution of HI at mildly non-linear scales. Based on the hydrodynamical simulations presented in \cite{Sarkar:2016lvb}, we approximate this bias using a two-dimensional mixed polynomial:
\begin{equation}
    b_{\mathrm{HI}}(k,z) = \sum_{m=0}^4\sum_{n=0}^2 c(m,n)k^mz^n\, .
\end{equation}
This analytical form remains valid for $k \lesssim 20 \, h/\mathrm{Mpc}$ within the redshift interval $z \sim 2-6$. While the bias is effectively constant at linear scales ($k \lesssim 0.1 \, \mathrm{Mpc}^{-1}$), it captures the complex stochasticity of HI clustering at smaller scales. The best-fit coefficients $c(m,n)$ are given by,
\begin{equation}
    c(m,n)\times10^2=\begin{pmatrix} 65.31 & 25.19 & 1.963 \\ -60.74 & 18.56 & 1.806 \\ 33.54 & -17.38 & 1.618 \\ -5.129 & 3.247 & -0.3803 \\ 0.2773 & -0.1899 & 0.02435 \end{pmatrix}\:,
\end{equation}
where rows correspond to powers of $k$ ($0 \leq m \leq 4$) and columns correspond to powers of $z$ ($0 \leq n \leq 2$).

To account for non-linear RSD, we implement a Lorentzian damping profile, $D_{\mathrm{FoG}} = (1 + k_{\parallel}^2 \sigma_{v,21}^2 / 2)^{-1}$. The pairwise velocity dispersion $\sigma_{v,21}$, which accounts for intra-halo kinematics, follows the redshift-dependent fitting function proposed in \cite{Sarkar:2018gcb}:
\begin{equation}
    \sigma_{v,21}(z) = \sigma_{v,0} (1 + z)^{-m} \exp\left[ - \left( \frac{z}{z_p} \right)^2 \right]\, ,
\end{equation}
with the optimized parameters $\sigma_{v,0}=11\,{\rm Mpc}$, $m=1.9$, and $z_p=11$. The  modelling of the HI signal helps us to resolve axionic suppressions in the matter power spectrum from astrophysical uncertainties in HI distribution.

\subsection{Lyman-\texorpdfstring{$\alpha$}{} power spectra}
\label{subsec:lymanalpha_ps}
The three-dimensional Ly$\alpha$ flux power spectrum serves as a biased tracer of the underlying total matter distribution. Within the mildly non-linear regime, the signal is modelled using the Kaiser approximation to account for linear RSD, combined with a phenomenological damping term that characterizes FoG effect.

The FoG suppression results from the stochastic, non-linear peculiar velocities of the IGM within virialized regions. This motion causes a kinematic smearing of the clustering signal along the line of sight, leading to a loss of power at high wavenumbers. Because ULAs primarily alter the matter power spectrum at small scales ($k \gtrsim 0.1 \, h \, \mathrm{Mpc}^{-1}$), a precise treatment of the FoG damping is essential to disentangle the cosmological axion signal from non-linear astrophysical velocity effects. By integrating these components, the 3D flux power spectrum is defined as
\begin{equation}
\label{eq:lya_ps}
    P_F^{\mathrm{3D}}(\boldsymbol{k},z) = b_F^2\left(1+\beta_F\mu^2\right)^2 P_{\mathrm{m}}(k,z) e^{-k^2 \mu^2 \sigma_{v,F}^2}\, ,
\end{equation}
where $\mu = k_{\parallel}/k$ and $\sigma_{v,F}$ denotes the line-of-sight velocity dispersion. We employ the redshift-dependent fitting functions for the flux bias $b_F$, the RSD parameter $\beta_F$ and the damping factor $\sigma_{v,F}$ derived from hydrodynamical simulations \cite{Arinyo-i-Prats:2015vqa, BOSS:2013rpr}.

The total observed Ly$\alpha$ power spectrum, $P_F^{\mathrm{T}}(\boldsymbol{k},z)$, is modelled as a sum of the signal and various noise contributions,
\begin{equation}
    P_F^{\mathrm{T}}(\boldsymbol{k},z) = P_F^{\mathrm{3D}}(\boldsymbol{k},z) + P_F^{\mathrm{1D}}(k_{\parallel},z)P_w^{\mathrm{2D}}(z) + P_N^{\mathrm{eff}}(z)\, .
\end{equation}
In this expression, the second term accounts for aliasing noise produced by the discrete sampling of quasar lines of sight, where $P_F^{\mathrm{1D}}$ represents the one-dimensional (1D) flux power spectrum and $P_w^{\mathrm{2D}}$ incorporates the transverse spatial weighting following \cite{McDonald:2006qs}. The final term, $P_N^{\mathrm{eff}}(z)$, represents the effective instrumental noise of the spectrograph for which we adopt the noise modelling according to our previous works~\cite{Pal:2026cgj,Bhaumik:2026mlc}.

\subsection{21-cm and  Lyman-\texorpdfstring{$\alpha$}{} cross power spectrum}
\label{subsec:cross_ps}
Extracting precise cosmological parameters from high-redshift auto-power spectra often necessitates marginalizing over intricate astrophysical biases and survey-specific foregrounds. The cross-correlation between the Ly$\alpha$ forest and 21-cm intensity mapping offers a novel method to mitigate these challenges. Since these two probes are observed through entirely independent channels, their respective instrumental systematics and primary foreground contaminants are expected to be statistically uncorrelated \cite{Sarkar:2019yea, Dash:2020yuq, Sarkar:2018vyh, Pal:2016icc, 2011MNRAS.410.1130G}.

Moreover, as these observables trace the DM distribution via distinct different bias mechanisms and physical environments, their cross-spectrum is highly effective at breaking degeneracies between non-linear astrophysical effects and the amplitude of matter fluctuations. This synergy makes the cross-correlation a remarkably clean tool for isolating the pristine cosmological signal. In the post-reionization era, the 21-cm and Ly$\alpha$ cross-spectrum is given by
\begin{equation}\label{eq:cross_spectra}
    P_{21,F}(\boldsymbol{k},z) = \bar{T}_{21}(z) b_F b_{\mathrm{HI}} \left(1+\beta_F\mu^2\right) \left(1+\beta_{\mathrm{HI}}\mu^2\right) P_{\mathrm{m}}(k,z) e^{-k^2 \mu^2 {\sigma_{v,F}^2}}\, ,
\end{equation}
where $P_{\mathrm{m}}(k,z)$ is the matter power spectrum, derived from \texttt{AxiCLASS}. In the absence of a dedicated model for the cross-correlation damping, we assume the FoG parameters to be consistent with those adopted for the Ly$\alpha$ auto-spectrum.

The ratio of the ULA model to the standard $\Lambda$CDM baseline for the cross-spectrum highlights the characteristic scale-dependent suppression signature. Due to the quantum pressure effects inherent in axionic dark matter, the cross-spectrum exhibits a significant reduction in power at high wavenumbers. Because the specific scale and profile of this suppression are exquisitely sensitive to the axion mass $m_a$ and its abundance, the cross-spectrum serves as a powerful diagnostic to break parameter degeneracies between axion physics and standard cosmological parameters.

\section{Detection prospects  of ULA dark matter}
\label{sec:detection_prospect}
Building on the theoretical framework for ULA DM in Sec.~\ref{sec:DMmodel}, we now turn to evaluating the observational viability of  ULA dark matter based on the instrumental specifications and the noise spectra of the missions of our consideration. For the spectroscopic optical tracing via the Ly$\alpha$ forest, we utilize a DESI-like spectrograph, covering a sky-area of 15,000~deg$^2$. For the 21-cm radio emissions, we adopt two distinct architectures: \textit{(i)} the SKA1-MID array operating in single-dish mode and \textit{(ii)} the proposed PUMA hexagonal close-packed interferometer. When evaluating the cross-correlation between the Ly$\alpha$ and 21-cm signals, the statistical power is fundamentally constrained by the overlapping sky coverage of the two distinct surveys. To remain conservative, we restrict the cross-spectrum volume to an overlap area of 2,000~deg$^2$. We adopt the noise modelling for the instrumental noise models and survey specifications from our previous works~\cite{Pal:2026cgj,Bhaumik:2026mlc}

A rigorous statistical framework is essential for accurately estimating the ULA parameters and disentangling their inherent degeneracies with the standard six cosmological parameters. In this section, we first quantify the detectability of the axionic suppression signatures by calculating the expected SNR across our considered observables. Following this SNR estimation, we present a comprehensive Fisher matrix analysis to forecast the projected joint constraints on the ULA parameter space. Ultimately, we aim to demonstrate how the synergy of these next-generation facilities can yield constraints on the ULA parameter space during the post-reionization epoch. Specifically, we investigate to provide constraints on the fraction of axion abundance for various axion masses ($10^{-30}~\text{eV}<m_a<10^{-20}~\text{eV}$). 

\subsection{Estimation of signal-to-noise ratio}
\label{subsec:SNR}
To quantify the detectability of the cosmological signal across different scales, we evaluate the SNR for a given tracer $i$. Within a specific redshift bin centred at $z_c$, the SNR can be calculated as
\begin{eqnarray}
\label{eq:SNR}
    \text{SNR}_i \equiv \sqrt{\frac{V_{\text{survey}}(z_c) k^3 \epsilon \, d\mu}{4\pi^2} \frac{P_i(k,z)^2}{\sigma^2[P_i(k,z)]}} \, ,
\end{eqnarray}
where $\epsilon \equiv \Delta k/k$ denotes the logarithmic bin width in $k$-space. The term $V_{\text{survey}}(z_c)$ represents the comoving survey volume contained within the redshift bin. This volume can be estimated by the fractional sky coverage of the instrument ($f_{\text{sky}}$) and the frequency bandwidth ($\Delta\nu$) as
\begin{eqnarray}
\label{eq:V_survey}
    V_{\text{survey}}(z_c) = \frac{4\pi}{3} f_{\text{sky}} \left[ \chi(z_{\text{max}})^3 - \chi(z_{\text{min}})^3 \right] \, ,
\end{eqnarray}
with $\chi(z_{\text{min}})$ and $\chi(z_{\text{max}})$ defining the comoving distances to the edges of the $z_c$-centred bin.

\begin{figure}[!ht]
    \centering
    \includegraphics[width=1\linewidth]{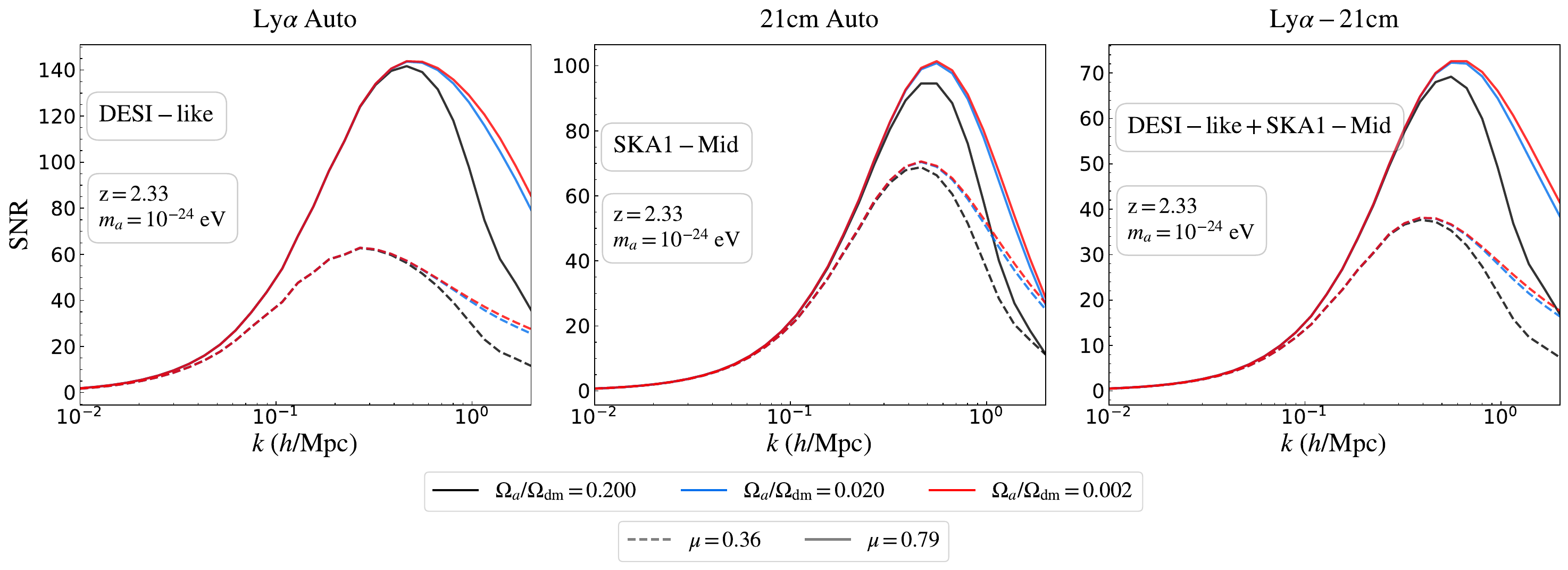}
    \caption{\textit{Illustration of SNR as a function of $k$ at $z=2.33$ for the Ly$\alpha$ auto-spectrum (\textbf{left panel}), the 21-cm auto-spectrum (\textbf{middle panel}) and the cross-spectrum (\textbf{right panel}), for various fractional abundances of ULAs, considering $m_a=10^{-24}$ eV. Ly$\alpha$ auto-spectrum assumes DESI-like surveys, while 21-cm auto-spectra are based on \textbf{SKA1-MID}. Solid and dashed lines correspond to different line-of-sight angles ($\mu$), highlighting the angular dependence of the signal.}}
    \label{fig:snr_ska}
\end{figure}

Fig.~\ref{fig:snr_ska} illustrates the estimated SNR as a function of wavenumber $k$ at $z=2.33$, for the Ly$\alpha$ auto-spectrum (left panel), the 21-cm auto-spectrum (middle panel) and their cross-spectrum (right panel). We explore various fractional abundances of ULAs for a fixed axion mass $m_a=10^{-24}$ eV. The figure depicts SNR estimation with the SKA1-MID for 21-cm auto-spectra. It is evident from the figure that SNR increases with the line-of-sight angle ($\mu$), as these modes closely align with the line-of-sight of the observations. However, at high-$k$, non-linear FoG effect significantly attenuates the SNR. On large scales (low $k$), the SNR closely tracks the standard CDM baseline, consistent with the fact that ULA dynamics mimic standard CDM at large scale (as seen in Fig.~\ref{fig:axion_PS}). The SNR peaks around $k\sim0.4$ (h/Mpc), where the ULA matter power spectrum exhibits its characteristic deviations from the $\Lambda$CDM model (\textit{cf.} Fig.~\ref{fig:axion_PS}). Furthermore, a larger ULA fraction increasingly suppresses the matter power spectrum, thereby decreasing the overall SNR. Assuming a foreground-free scenario and $\mu=0.79$, the Ly$\alpha$ auto-spectrum yields the highest sensitivity, reaching a maximum SNR of approximately 141 for $\Omega_a/\Omega_{\rm dm}=0.2$. The SKA1-MID 21-cm auto-spectrum reaches a peak SNR of $\sim95$, while the lowest SNR-peak exhibits for cross-correlation which is $\sim70$~\footnote{Note that this relative hierarchy in SNR peaks may shift depending on the specific foreground mitigation strategies applied to the observational data. While in the case of cross-spectra, this also depends on the overlapped sky coverage area considered.}.

\begin{figure}[!ht]
    \centering
    \includegraphics[width=1\linewidth]{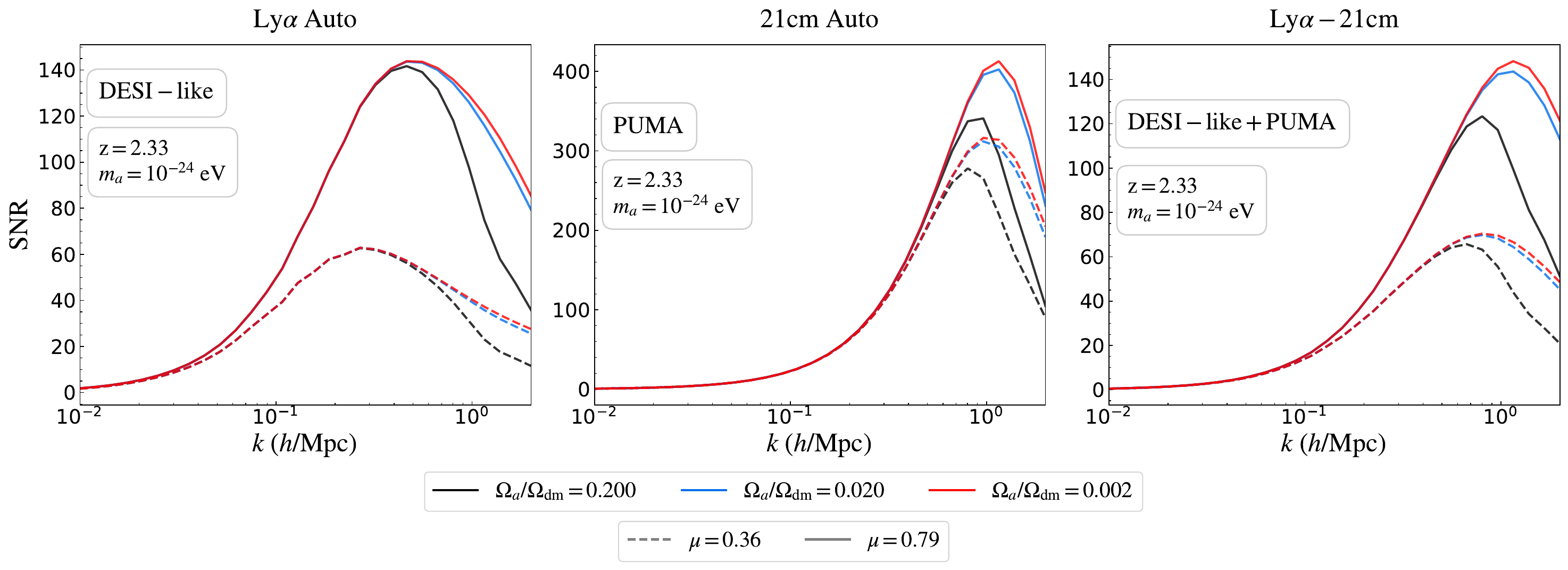}
    \caption{\textit{Similar representation as Fig.~\ref{fig:snr_ska}, but this time for \textbf{PUMA} instead of SKA1-MID.}}
    \label{fig:snr_puma}
\end{figure}

Fig.~\ref{fig:snr_puma} illustrates the SNR as a function of wavenumber $k$, mirroring the analysis in Fig.~\ref{fig:snr_ska} but substituting the SKA1-MID configuration with PUMA~\footnote{For detailed noise configurations of PUMA (\href{https://www.puma.bnl.gov/}{https://www.puma.bnl.gov/}) and SKA1-MID (\href{https://www.skao.int/}{https://www.skao.int/}), we refer the reader to Section 4 of~\cite{Pal:2026cgj}.}. Although the fundamental scale-dependent profile of the SNR remains unchanged, PUMA yields a substantially higher sensitivity owing to its advanced instrumental specifications. For instance, at $z=2.33$, the peak SNR for 21-cm auto-spectrum reaches roughly 95 for SKA1-MID; in contrast, PUMA achieves a peak SNR of $\sim 340$ for an axion mass of $m_a = 10^{-24}$ eV and a fractional abundance of $\Omega_a/\Omega_{\rm dm} = 0.2$. Consequently, evaluating this identical ULA parameter space with PUMA enhances the peak SNR of the cross-spectrum to approximately 125. 

\begin{figure}[!ht]
    \centering
    \includegraphics[trim ={0 3.2cm 0 0}, clip, width=1\linewidth]{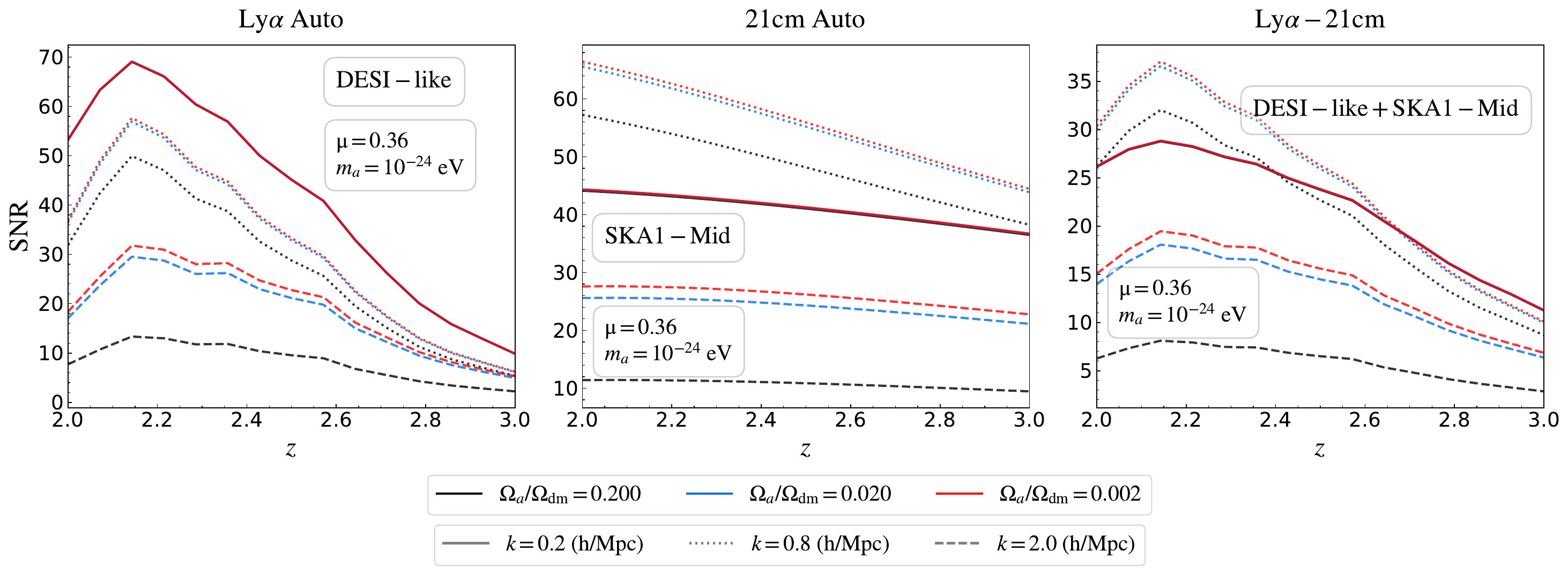}
    \includegraphics[width=1\linewidth]{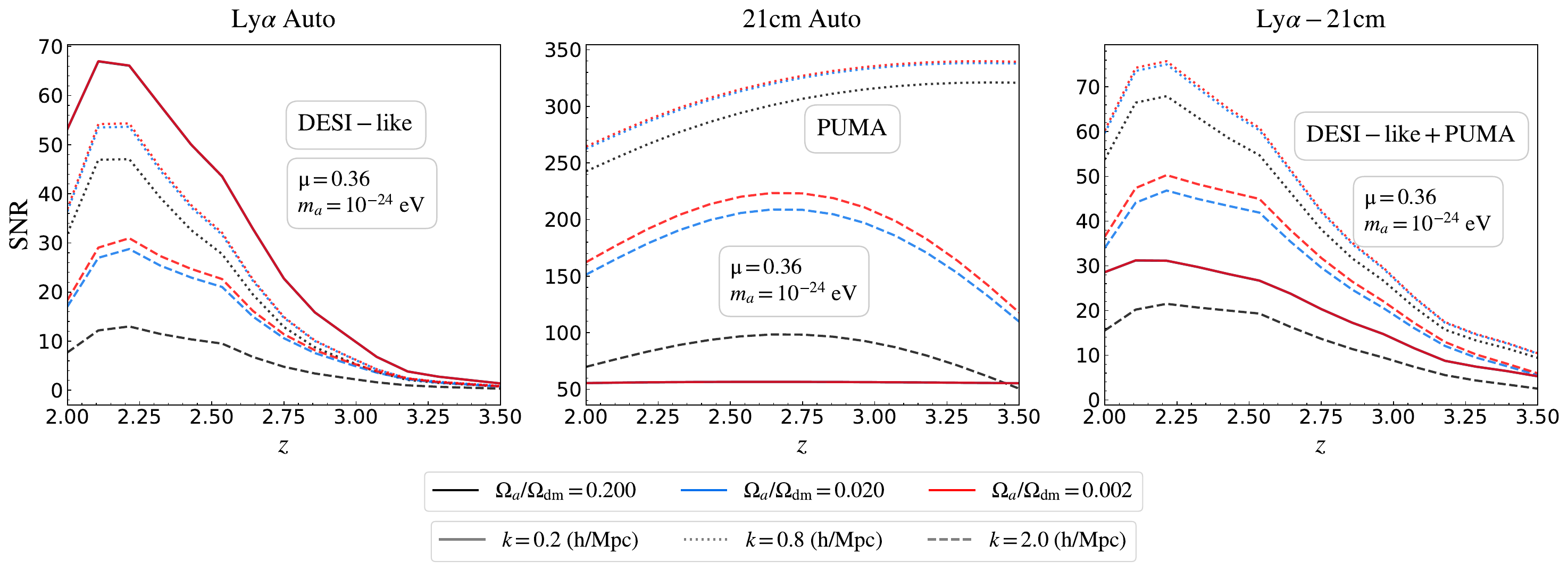}
    \caption{\textit{Illustration of SNR as a function of $z$ for various wavenumbers $k$ and fractional ULA abundances, assuming a fixed axion mass of $m_a=10^{-24}$~eV and a line-of-sight angle of $\mu=0.36$. In both rows, the \textbf{left panels} display the Ly$\alpha$ auto-spectrum (DESI-like), the \textbf{middle panels} show the 21-cm auto-spectrum, and the \textbf{right panels} present the cross-spectrum. The \textbf{top row} utilizes SKA1-MID for the 21-cm observations, while the \textbf{bottom row} utilizes the PUMA interferometer.}}
    \label{fig:snr_z_profile}
\end{figure}

 Now, it is the time to turn to the redshift evolution of the SNR, presented in Fig.~\ref{fig:snr_z_profile}. Our analysis is restricted to the window $z \in [2, 3]$ for SKA1-MID and $z \in [2, 3.5]$ for PUMA respectively, which represents the optimal overlapping epoch for these specific Ly$\alpha$ and 21-cm observations. The top row illustrates the Ly$\alpha$ auto-spectra, 21-cm auto-spectra (using SKA1-MID), and their cross-spectra from left to right, assuming $\mu=0.36$ and $m_a=10^{-24}$~eV across several benchmark values for $\Omega_a/\Omega_{\rm dm}$ and $k$. For the Ly$\alpha$ and cross-spectra, the sensitivity peaks near $z \sim 2.1$. At higher redshifts, the increasingly dispersed distribution of quasars suppresses the measurable power spectra, naturally attenuating the SNR. However, the 21-cm auto-spectrum for SKA1-MID exhibits a monotonic decrease across this entire redshift range. Furthermore, for a given scale $k$, the SNR drops as the axion fractional abundance increases. This occurs because a larger ULA fraction induces a more pronounced suppression of the small scale matter power spectrum. These small scale suppressions are most effectively captured by Ly$\alpha$ experiments like DESI. While the high-$k$ modes yield a lower SNR due to this inherent physical suppression, they are uniquely vital for distinguishing between different ULA fractions. For example, at $k=2.0~h/{\rm Mpc}$, the DESI-like survey yields a peak-SNR of $\sim 14$ for $\Omega_a/\Omega_{\rm dm}=0.2$,  which improves to $\sim 30$ for $\Omega_a/\Omega_{\rm dm}=0.002$. Conversely, on larger scales ($k=0.2~h/{\rm Mpc}$) where the ULA model aligns almost perfectly with the standard CDM prediction, the peak SNR reaches an impressive $\sim 68$, but the curves for different axion fractions become entirely indistinguishable. It is also important to note that the SNR scaling with $k$ is non-linear for 21-cm observations due to the strong influence of non-linear RSD effects. Consequently, SKA1-MID achieves a higher SNR at $k=0.8~h/{\rm Mpc}$ than at $k=2.0~h/{\rm Mpc}$. This same hierarchical trend is mirrored in the cross-spectrum, which achieves a peak SNR of roughly 30 for $\Omega_a/\Omega_{\rm dm}=0.2$ when combining the DESI-like and SKA1-MID surveys.

The bottom row of Fig.~\ref{fig:snr_z_profile} depicts similar analysis, by substituting SKA1-MID with the PUMA interferometer for the 21-cm observations. Unlike the monotonic decrease seen with SKA1-MID, PUMA's 21-cm auto-spectrum SNR reveals a distinct peak around $z \sim 2.75$ to $3.25$ for all considered wavenumbers before attenuating at higher redshifts. Moreover, PUMA's advanced instrumental specifications drastically amplify the sensitivity, pushing the peak 21-cm auto-spectrum SNR up to $\sim 350$ for $k=0.8~h/{\rm Mpc}$. Consequently, the synergistic combination of PUMA and a DESI-like survey elevates the cross-spectrum peak SNR to $\sim 65$ (for $m_a=10^{-24}$~eV, $\Omega_a/\Omega_{\rm dm}=0.2$, and $k=0.8~h/{\rm Mpc}$).

While this SNR analysis clearly highlights the enhanced detection prospects by the next-generation arrays like PUMA, the estimation of the  uncertainties in the parameters of our concern and the complex covariance between them looks quite promising.  
This can be materialized with the help of the above SNR plane, by picking fiducial values of the parameters from high SNR regions followed by  a Fisher matrix analysis around that fiducial value, 
leading to an investigation of allowed regions of high SNR and low error.
We detail that in the following section.

\subsection{Fisher matrix forecast on the parameters}
\label{subsec:Fisher}

\subsubsection{Construction of Fisher matrix and covariance matrix}
\label{subsubsec:Fisher_setup}

In order to estimate the precision level that can be achieved in upcoming observations for the relevant parameters of ULA,  we have made use of the Fisher information matrix formalism~\cite{Tegmark:1996bz,2009arXiv0906.0664H,Bharadwaj:2015vwa}. The Fisher matrix analysis estimates the uncertainties on the ULA parameters  from the LSS missions and the CMB-S4-like missions as representative for future stage 4 CMB missions. Assuming a Gaussian likelihood $\mathcal{L}$, the elements of the Fisher matrix are defined by the negative expectation over observations of the second order derivative of the log-likelihood
\begin{equation}
    F_{\alpha\beta} \equiv -\left\langle \frac{\partial^2 \ln \mathcal{L}}{\partial \theta_\alpha \, \partial \theta_\beta} \right\rangle.
\end{equation}
For a parameter vector $\boldsymbol{\theta}$ containing $n$ fiducial parameters, the corresponding $(n\times n)$-dimensional Fisher matrix for the $i^{\mathrm{th}}$ observable is formulated as
\begin{equation}
    F_{\alpha\beta}^{(i)}(\boldsymbol{\theta})=\sum\limits_{m}^{\rm k-bins}\sum\limits_{n}^{\rm z-bins}\sum_{p}^{\mu{\rm -bins}}\frac{1}{\sigma^2\left[P_i\left(k_m,z_n,\mu_p\right)\right]}\frac{\partial P_i\left(k_m,z_n,\mu_p\right)}{\partial\theta_\alpha}\frac{\partial P_i\left(k_m,z_n,\mu_p\right)}{\partial\theta_\beta},
\end{equation}
where the indices $\alpha$ and $\beta$ span the parameter space. Inverse of this matrix gives the covariance matrix. The diagonal elements of this inverted matrix provide the projected $1\sigma$ marginal errors, while the off-diagonal components quantify the correlations between parameters.

To thoroughly evaluate the physical degeneracies between the ULA characteristics and standard cosmological metrics, we define our complete parameter vector $\boldsymbol{\theta}$ as
\begin{equation}
    \boldsymbol{\theta} \equiv \left\{ \omega_b, \Omega_{\rm dm}, h, \tau_{\rm reio}, \ln(10^{10}A_s), n_s, m_a, \Omega_a/\Omega_{\rm dm} \right\}.
\end{equation}
Given that primary CMB observations and post-reionization LSS surveys probe distinct cosmic volumes and structural regimes, their respective datasets can be treated as statistically independent. Consequently, the total parameter covariance matrix is computed by inverting the sum of the individual Fisher matrices
\begin{equation}
    \mathbf{C}_{\alpha\beta} \equiv \left( F^{\rm LSS} + F^{\rm CMB} \right)^{-1}_{\alpha\beta}.
\end{equation}
The marginalised $1\sigma$ uncertainty for any specific parameter $\theta_\alpha$ can be evaluated simply as $\sqrt{\mathbf{C}_{\alpha\alpha}}$.

\subsubsection{Results and analysis}
\label{subsubsec:Fisher_analysis}
We now present the results of our Fisher matrix forecasts, quantifying the efficiency of upcoming multi-tracer surveys to constrain the ULA parameter space. Our analysis is based on a fiducial choice of the standard cosmological parameters and the ULA fraction as shown in Table~\ref{tab:fiducial_parameters}. The fiducial choice for cosmological parameters are fixed at the best-fit values of Planck-18 TT,TE,EE+lowP+lensing+BAO~\cite{Planck:2018vyg}, while the fiducial for $\Omega_a/\Omega_{\rm dm}$ is motivated from Refs.~\cite{Hlozek:2017zzf,Bauer:2020zsj}. The Fisher matrix analysis is performed for the set of fiducial parameters across the axion mass range $m_a\in [10^{-30}, 10^{-20}]$ eV, with unit logarithm spacing to obtain the estimated uncertainties on $\Omega_a/\Omega_{\rm dm}$ for different ULA masses. Throughout the analysis, we adopted 10 logarithm bins of $k$-space in the range $[0.1,1]$ h/Mpc and 5 equally spaced $\mu$ bins in the range $[0, 1]$. While 5 equally spaced redshift bins in the range of $[2, 3.05]$ are considered for SKA1-MID, an extended redshift range of $[2, 3.5]$ is adopted for PUMA, keeping same redshift binning.

\begin{table}[!ht]
    \centering
    \renewcommand{\arraystretch}{1.3}
    \begin{tabular}{l| c c c c c c c}
        \hline\hline
        Parameter & $\omega_b$ & $\omega_{\rm cdm}$ & $h$ & $\tau$ & $\ln(10^{10}A_s)$ & $n_s$ & $\Omega_a/\Omega_{\rm dm}$ \\
        \hline
        Fiducial value & 0.02242 & 0.11933 & 0.6766 & 0.0561 & 3.047 & 0.9665 & 0.02 \\
        \hline\hline
    \end{tabular}
    \caption{\textit{Fiducial values of the cosmological and ULA parameters used for the Fisher matrix analysis. The standard cosmological parameters are set to the best-fit values from Planck-18 TT,TE,EE+lowP+lensing+BAO~\cite{Planck:2018vyg}, while the fiducial value for $\Omega_a/\Omega_{\rm dm}$ is motivated by Refs.~\cite{Hlozek:2017zzf,Bauer:2020zsj}.}}
    \label{tab:fiducial_parameters}
\end{table}

\begin{figure}[!ht]
    \centering
    \includegraphics[width=0.9\linewidth]{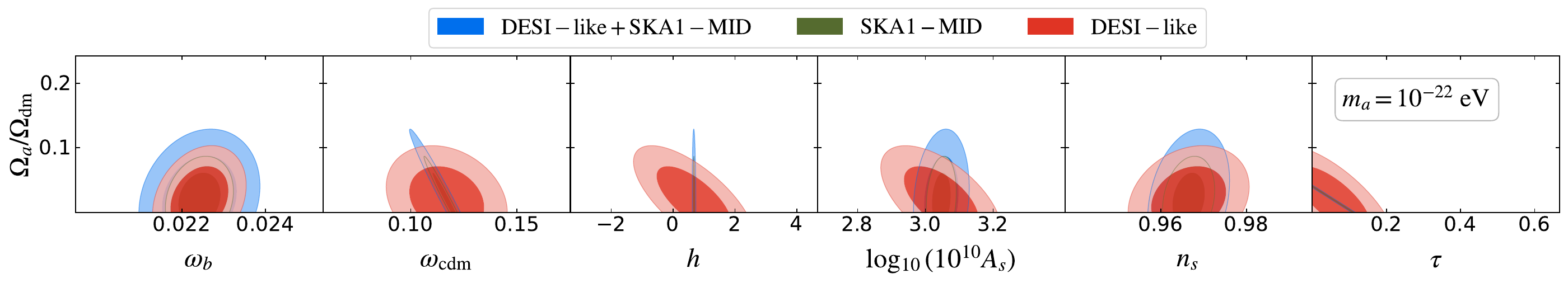}
    \includegraphics[trim= 0 0 0 1.5cm, clip, width=0.9\linewidth]{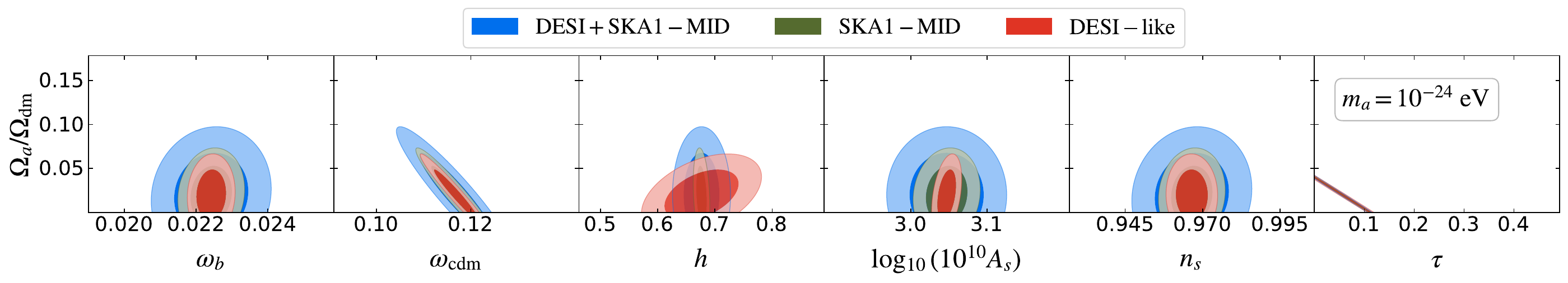}
    \includegraphics[trim= 0 0 0 1.5cm, clip, width=0.9\linewidth]{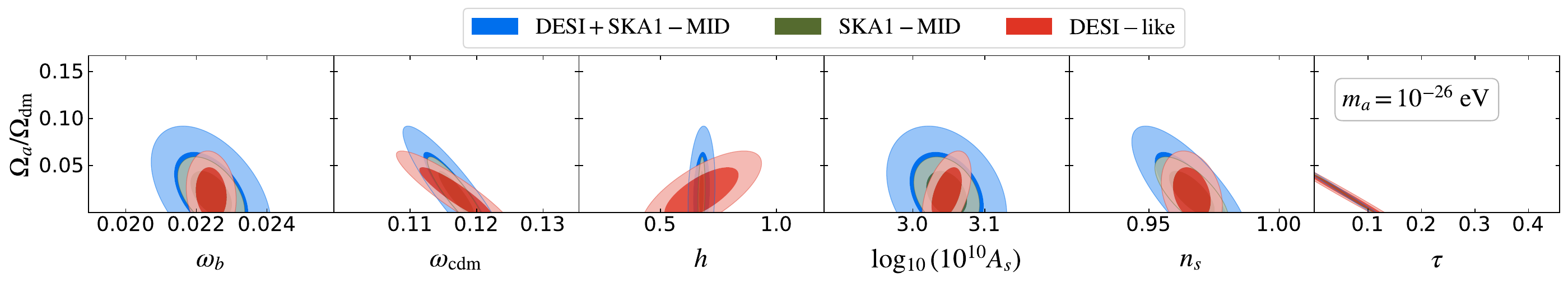}
    \includegraphics[trim= 0 0 0 1.5cm, clip, width=0.9\linewidth]{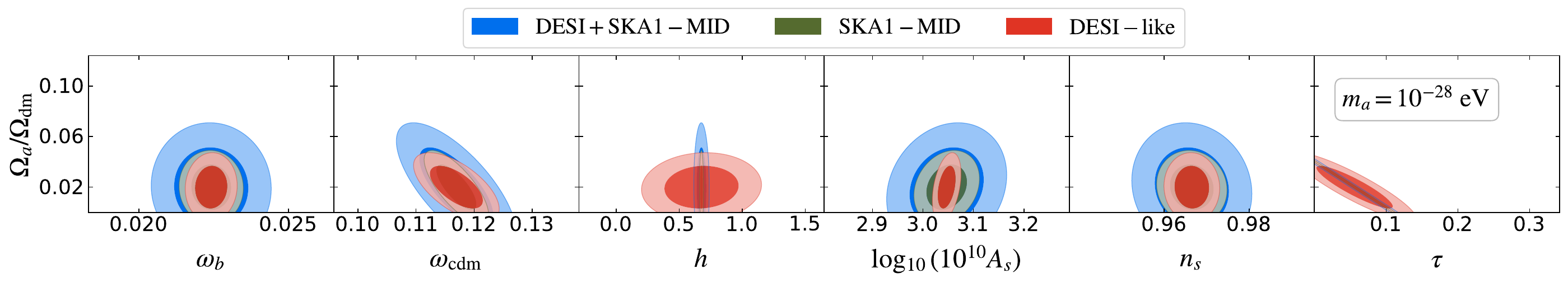}
    \caption{\textit{Marginalised 2D Fisher contours of fractional abundance of ULAs with other cosmological parameters for four different $m_a$. \textbf{SKA1-MID} is utilized for 21-cm observations. Each row presents the $1\sigma$ and $2\sigma$ contours for different axion masses.}}
    \label{fig:fisher_2d_ska}
\end{figure}

Fig.~\ref{fig:fisher_2d_ska} presents marginalised 2D  contours  for Ly$\alpha$ auto-spectrum, 21-cm auto-spectrum and their cross-spectrum from Fisher analysis. By taking into account DESI-like survey as Ly$\alpha$ observation and SKA1-MID as 21-cm observation, the figure depicts the correlation between the fractional abundance of ULAs and the standard six cosmological parameters for four different ULA masses. The fractional abundance $\Omega_a/\Omega_{\rm dm}$ maintains a negative correlation with $\omega_{\rm cdm}$ throughout all the ULA mass ranges considered. This correlation is expected, as the total dark matter abundance is strictly the sum of the abundances of CDM and ULA components. Thus, 
in order to maintain the global clustering amplitude,
an increase in the axion fraction results in a corresponding decrease in $\omega_{\rm cdm}$. Interestingly, the correlation with the baryon density ($\omega_b$) behaves differently depending on the tracer. For the Ly$\alpha$ auto-spectrum, $\Omega_a/\Omega_{\rm dm}$ is largely uncorrelated with $\omega_b$, as the Ly$\alpha$ flux primarily traces the total matter power spectrum. Conversely, the 21-cm auto-spectrum explicitly depends on the neutral hydrogen abundance, introducing a distinct correlation between $\omega_b$ and the axion fraction. Our analysis reveals that this correlation is slightly positive for $m_a \gtrsim 10^{-25}$~eV, but gets inverted for lighter masses ($m_a \lesssim 10^{-25}$~eV), establishing $m_a \sim 10^{-25}$~eV as a phenomenological turning point. 
Because the 21-cm observations dominate the parameter covariance, the cross-correlation contours naturally follow the degeneracy directions dictated by SKA1-MID.

\begin{figure}[!ht]
    \centering
    \includegraphics[width=0.9\linewidth]{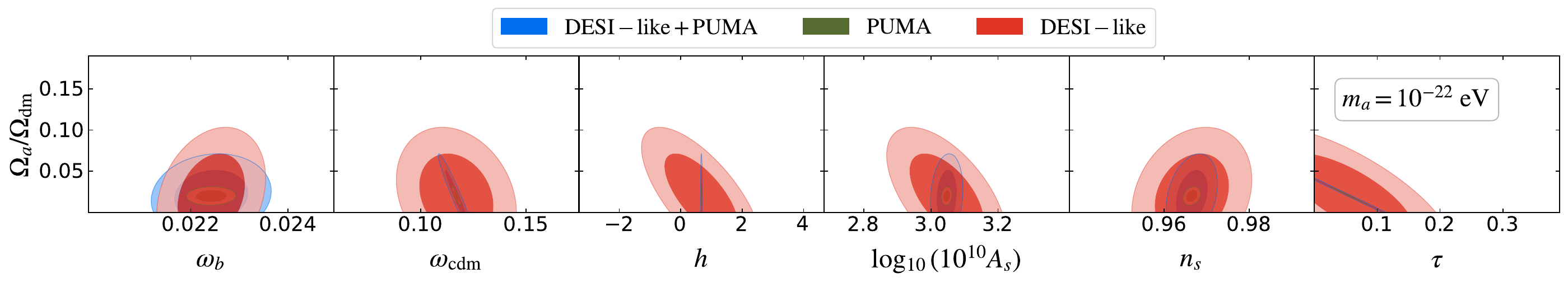}
    \includegraphics[trim= 0 0 0 1.5cm, clip, width=0.9\linewidth]{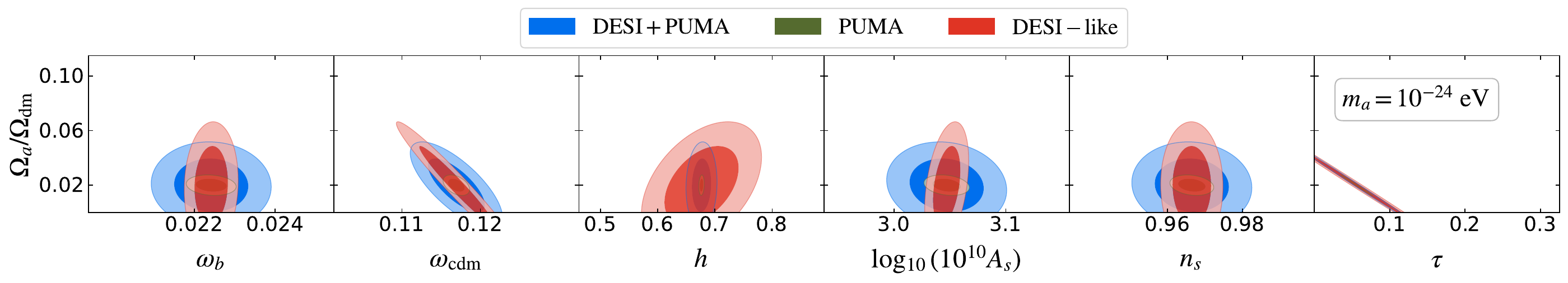}
    \includegraphics[trim= 0 0 0 1.5cm, clip, width=0.9\linewidth]{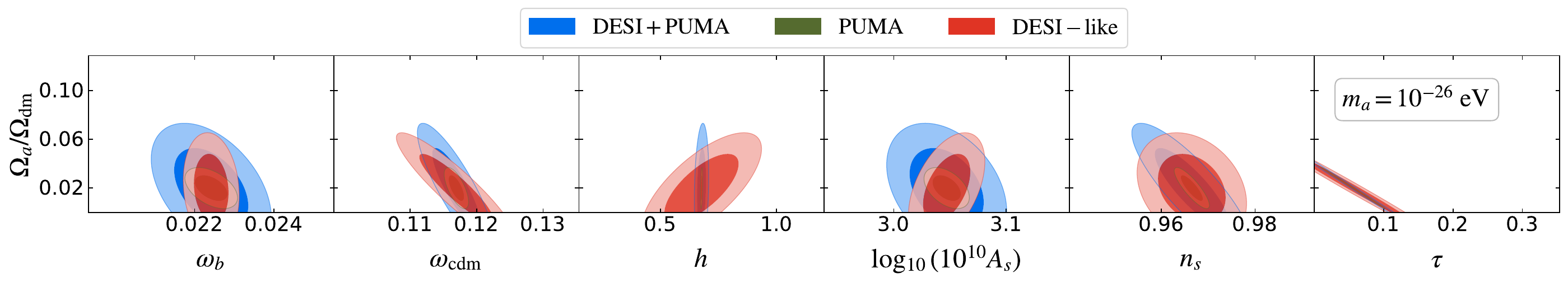}
    \includegraphics[trim= 0 0 0 1.5cm, clip, width=0.9\linewidth]{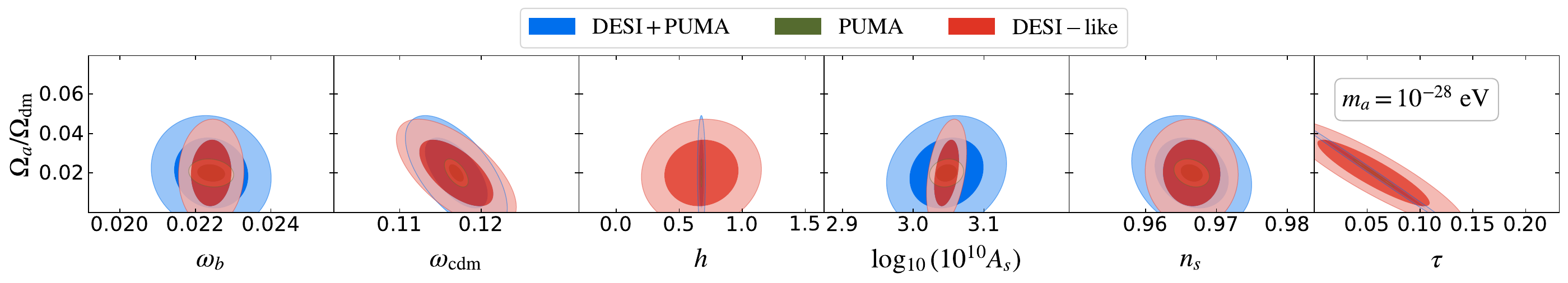}
    \caption{\textit{Same level of information as in Fig.~\ref{fig:fisher_2d_ska} but this time \textbf{PUMA} instead of SKA1-MID.}}
    \label{fig:fisher_2d_puma}
\end{figure}

We extend this analysis to the PUMA interferometer, with the corresponding  contours   presented in Fig.~\ref{fig:fisher_2d_puma}. Due to PUMA's improved specification, the estimated errors on $\Omega_a/\Omega_{\rm dm}$ are further narrower down compared to SKA1-MID. While the correlation angle between $\omega_b$ and $\Omega_a/\Omega_{\rm dm}$ depends on ULA mass, similar to SKA1-MID, the correlation-flip shifts towards a higher mass of $m_a \sim 10^{-24}$ eV. The correlation between $\omega_{\rm cdm}$ and $\Omega_a/\Omega_{\rm dm}$ remains intact.

\begin{figure}[!ht]
    \centering
    \includegraphics[width=0.4\linewidth]{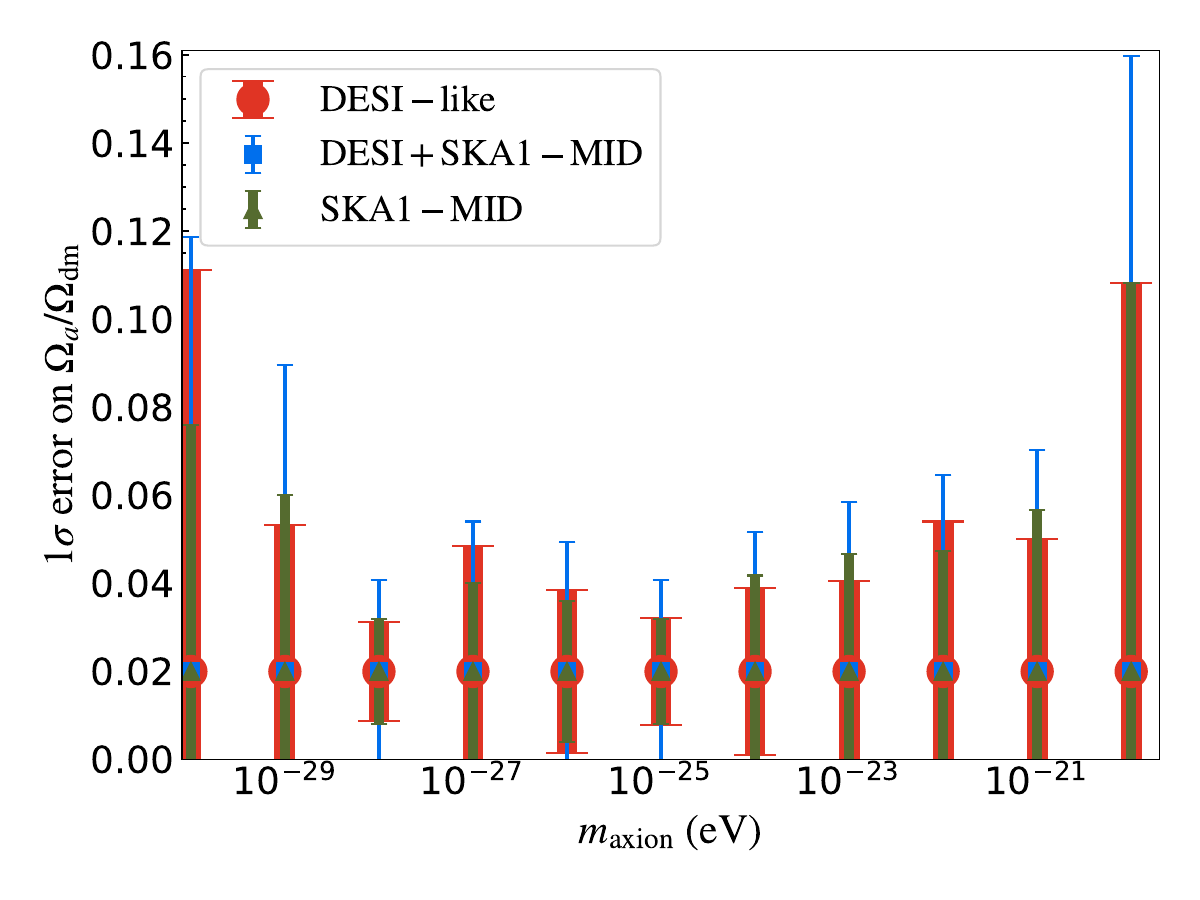}
    \includegraphics[width=0.4\linewidth]{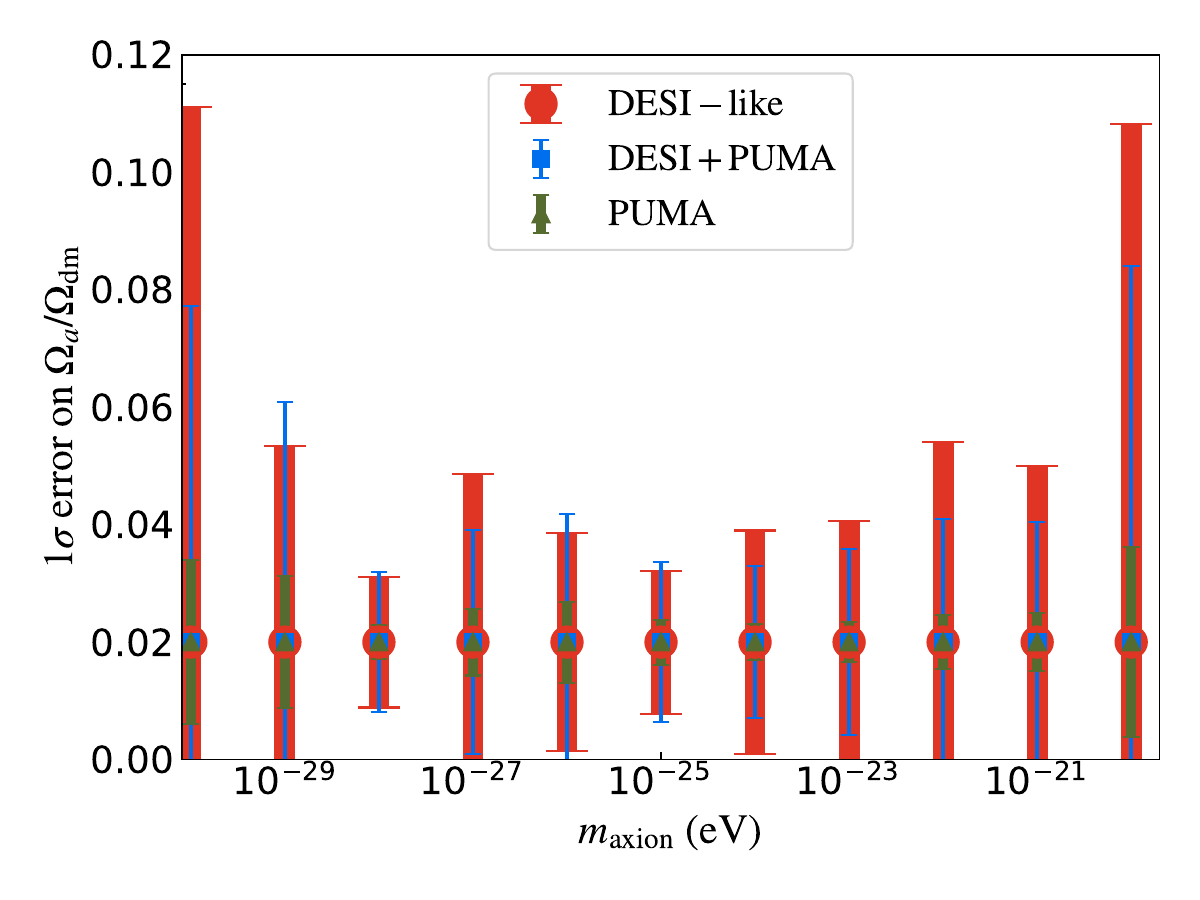}
    \caption{\textit{Presenting 1$\sigma$ error-bar on $\Omega_a/\Omega_{\rm dm}$ for several ULA masses, while the fiducial value for $\Omega_a/\Omega_{\rm dm}$ is set to $0.02$. Each panel presents the projected constraints on $\Omega_a/\Omega_{\rm dm}$ from Ly$\alpha$ and 21-cm auto-spectra and their cross-spectra. \textbf{Left panel} presents SKA1-MID for 21-cm observations, while \textbf{right panel} shows the results for PUMA. Both the panels depicts that the fractional abundance of ULAs can be constrained tightly around $m_a\sim 10^{-25}$ eV. PUMA appears as a promising interferometric setup to probe $\Omega_a/\Omega_{\rm dm}$.}}
    \label{fig:1sigma_omega_scf_wo_cmb}
\end{figure}

Fig.~\ref{fig:1sigma_omega_scf_wo_cmb} summarizes these results across the full ULA mass range, showing the projected $1\sigma$ marginal uncertainties on $\Omega_a/\Omega_{\rm dm}$ as a function of $m_a$, while the fiducial value for $\Omega_a/\Omega_{\rm dm}$ is set to $0.02$.
Under the optimistic assumption of foreground-free scenario, the individual 21-cm auto-spectra provide better constraints than the Ly$\alpha$ auto-spectrum, with PUMA emerging as the most sensitive configuration. However, auto-spectra are subjected to residual foregrounds and instrumental systematics.
Although the estimated errors on $\Omega_a/\Omega_{\rm dm}$ for the cross-spectrum is larger than the auto-spectra, the projected $1\sigma$ uncertainties for cross-spectra will be more realistic as cross-spectrum effectively isolates the shared cosmological signal while actively mitigating independent survey systematics~\footnote{The rigorous modelling and removal of foregrounds remain a primary challenge for 21-cm cosmology. Since the cross-correlation inherently suppresses these uncorrelated contaminants, it serves as the focal point of our analysis. Detailed foreground mitigation strategies for the auto-spectra are left for future dedicated studies.}.  Among the cross-correlation setups, the synergy of a DESI-like survey with PUMA estimates small uncertainties in probing $\Omega_a/\Omega_{\rm dm}$. 
The estimated errors on fractional ULA abundance reveal an optimal sensitivity window around $m_a \sim 10^{-25}$~eV. For heavier axions ($m_a \gtrsim 10^{-22}$~eV), the characteristic Jeans scale ($k_J$) shifts to highly non-linear regimes. As a result, the ULA fluid becomes indistinguishable from CDM on the target scales ($k \in [0.1, 1]~h/{\rm Mpc}$), washing out the signature of $\Omega_a/\Omega_{\rm dm}$ and increasing the uncertainties in measurement. Conversely, for $m_a \lesssim 10^{-28}$~eV, the Jeans scale shifts to linear regimes (low-$k$). Thus, the axions do not help in clustering within the observed $k$-window, resulting in a deficit in the total matter density. This leads to a significant correlation between $\omega_{\rm cdm}$ and $\Omega_a/\Omega_{\rm dm}$ (as seen in the bottom rows of Figs.~\ref{fig:fisher_2d_ska} and \ref{fig:fisher_2d_puma}), substantially increasing the marginal errors.

While the cross-correlations between Ly$\alpha$ and 21-cm observations appear as a potential probe for ULA-parameters, the estimated uncertainties on the other cosmological parameters, specially on $\tau$, turn out to be quite unimpressive.  On the other hand, CMB experiments can provide tighter constraints on the cosmological parameters~\cite{Planck:2018vyg}. Furthermore, the impact of ULA parameters on the CMB TT spectrum (see Fig.~\ref{fig:axion_CMB}) shows the potential to estimate the ULA parameters using upcoming CMB experiments. Thus, a joint analysis of future CMB experiment with 21-cm and Ly$\alpha$ experiments may estimate the smaller errors on the fraction of ULA-abundance for different axion masses. With this in mind, we utilize CMB-S4-like stage 4 CMB experimental setup to estimate the uncertainties on ULA parameters. 

\begin{figure}[!ht]
    \centering
    \includegraphics[width=0.9\linewidth]{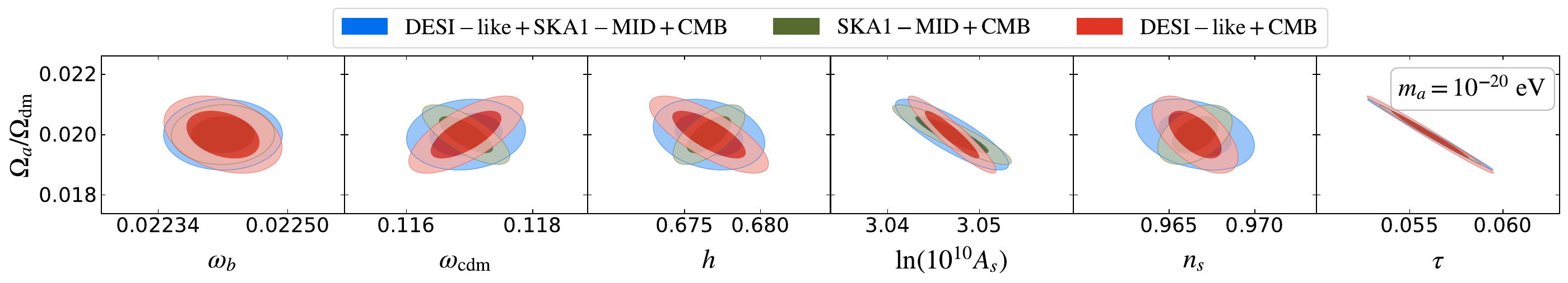}
    \includegraphics[trim= 0 0 0 1.5cm, clip, width=0.9\linewidth]{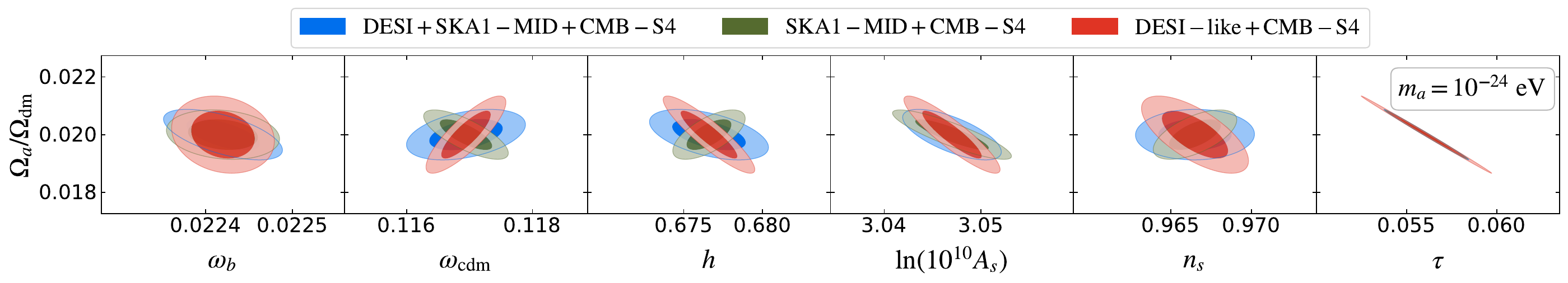}
    \includegraphics[trim= 0 0 0 1.5cm, clip, width=0.9\linewidth]{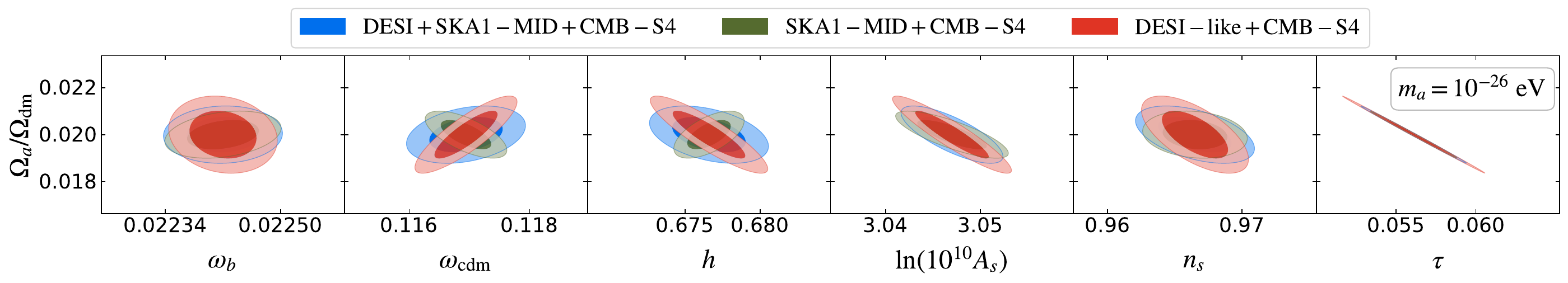}
    \includegraphics[trim= 0 0 0 1.5cm, clip, width=0.9\linewidth]{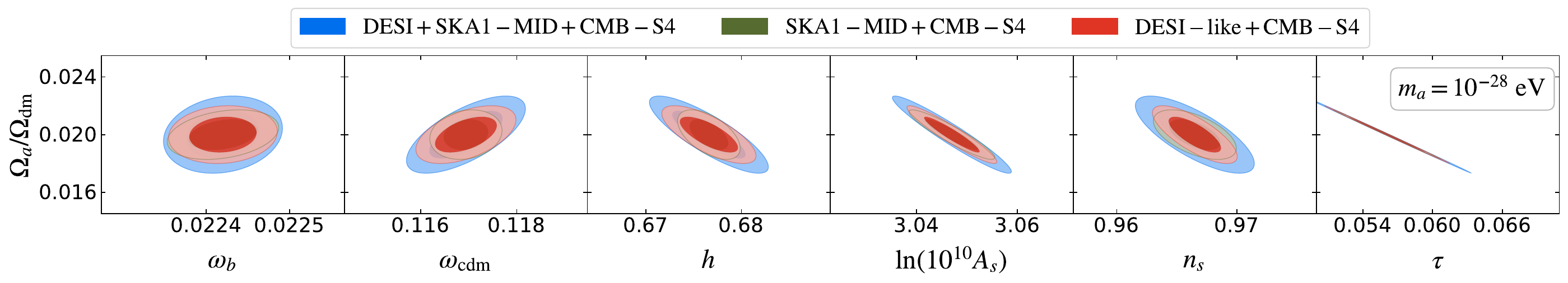}
    \caption{\textit{Marginalised 2D Fisher contour of fractional abundance of ULAs with other cosmological parameters for four different $m_a$, considering joint analysis with CMB-S4-like experimental setup. \textbf{SKA1-MID} is utilized for 21-cm observations. Each row presents the 1$\sigma$ and $2\sigma$ contours for different axion masses.}}
    \label{fig:fisher_2d_cmb_ska}
\end{figure}

\begin{figure}[!ht]
    \centering
    \includegraphics[width=0.9\linewidth]{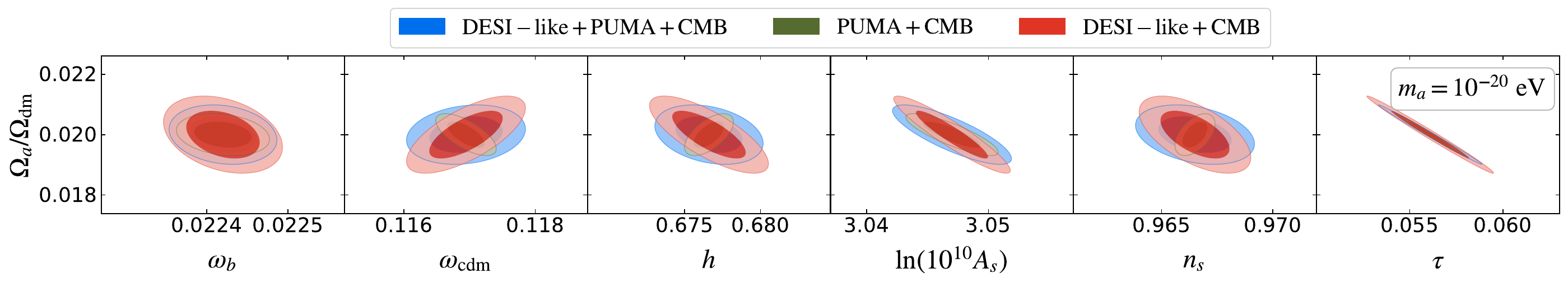}
    \includegraphics[trim= 0 0 0 1.5cm, clip, width=0.9\linewidth]{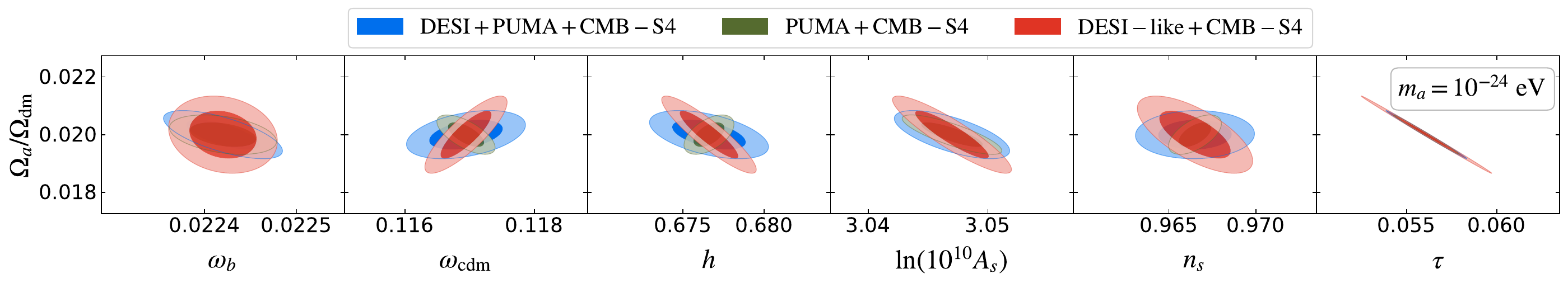}
    \includegraphics[trim= 0 0 0 1.5cm, clip, width=0.9\linewidth]{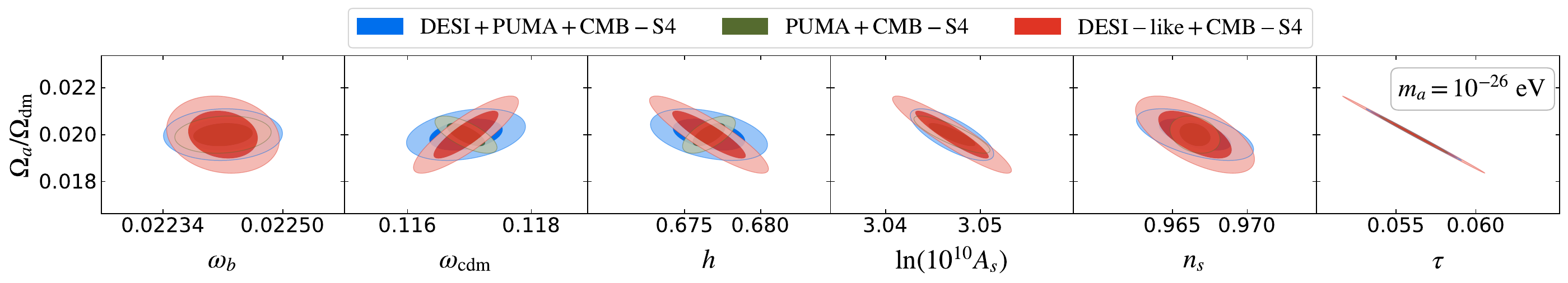}
    \includegraphics[trim= 0 0 0 1.5cm, clip, width=0.9\linewidth]{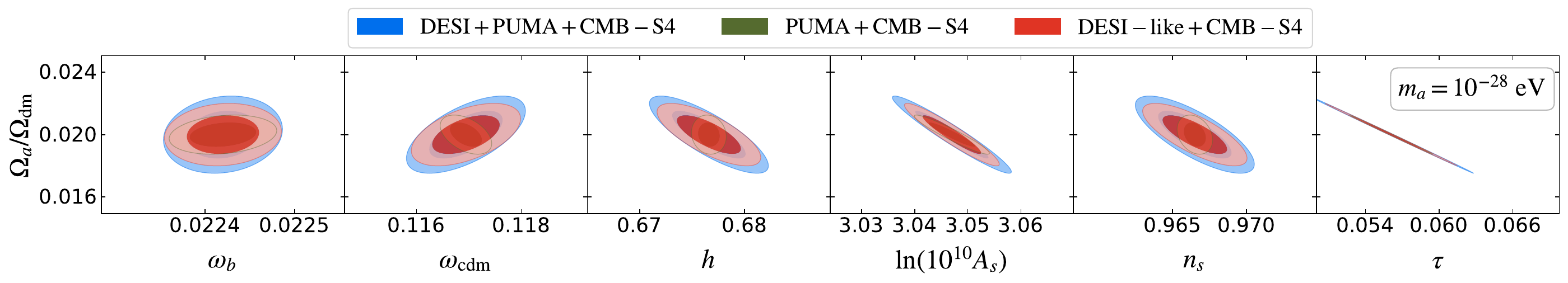}
    \caption{\textit{Same representation as Fig.~\ref{fig:fisher_2d_cmb_ska}, but utilising the \textbf{PUMA} interferometer for the 21-cm observations instead of SKA1-MID.}}
    \label{fig:fisher_2d_cmb_puma}
\end{figure}

Fig.~\ref{fig:fisher_2d_cmb_ska} presents the 2D marginalised contours resulting from the joint Fisher analysis of transcending experiments similar to CMB-S4 with the Ly$\alpha$ auto-spectrum, the 21-cm auto-spectrum (using SKA1-MID) and their cross-spectrum. The inclusion of CMB-S4-like experiments result in a reasonable estimation of the uncertainties on the ULA fraction, that also reflect on the uncertainties of the six cosmological parameters  compared to the LSS-only configurations. Fig.~\ref{fig:fisher_2d_cmb_puma} shows the corresponding results when SKA1-Mid is replaced by the more sensitive PUMA interferometer.
An interesting feature in both figures is the correlation between $\omega_{\rm cdm}$ and the ULA fraction. As discussed previously, LSS-only configurations are primarily sensitive to the total dark matter density $\omega_{\rm dm}$ via the clustering amplitude and the redshift of matter-radiation equality. Thus, a significant negative correlation between $\omega_{\rm cdm}$ and the ULA fraction ULA  is naturally enforced by keeping $\omega_{\rm dm}$ nearly constant.
However, as expected, CMB-S4-like future experiments act as a good probe for the primordial parameters ($A_s$, $n_s$) and the reionization optical depth ($\tau$), unlike the Ly$\alpha$ forest.  While the isolated Ly$\alpha$ and 21-cm data leave the ULA fraction largely uncorrelated with these primordial parameters, 
the joint CMB-S4-like+DESI-like+21-cm analysis securely anchors them. This synergistic approach demonstrates the potential to estimate the errors simultaneously on ULA parameters and primordial parameters, and a positive correlation between $\omega_{\rm cdm}$ and fractional abundance of ULA dark matter.

\begin{figure}[!ht]
    \centering
    \includegraphics[width=0.4\linewidth]{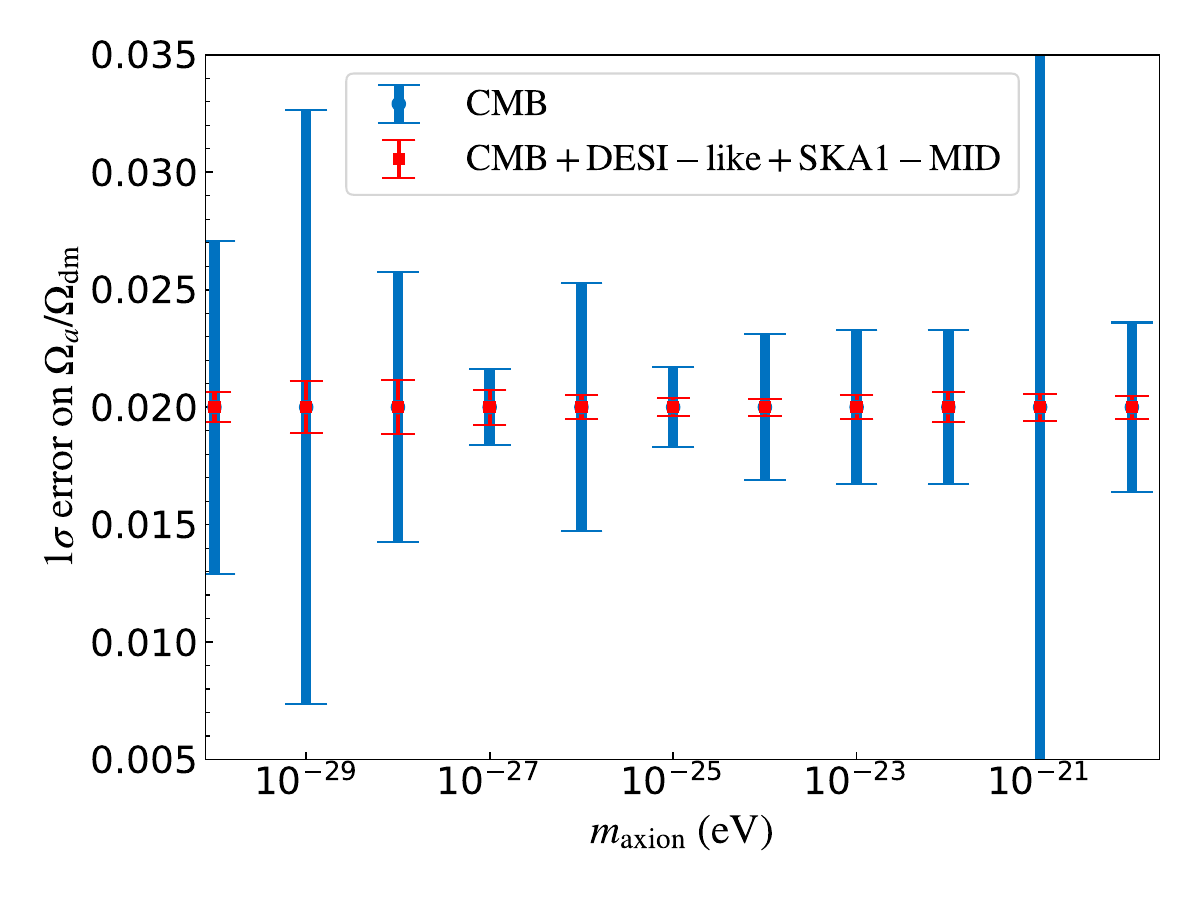}
    \includegraphics[width=0.4\linewidth]{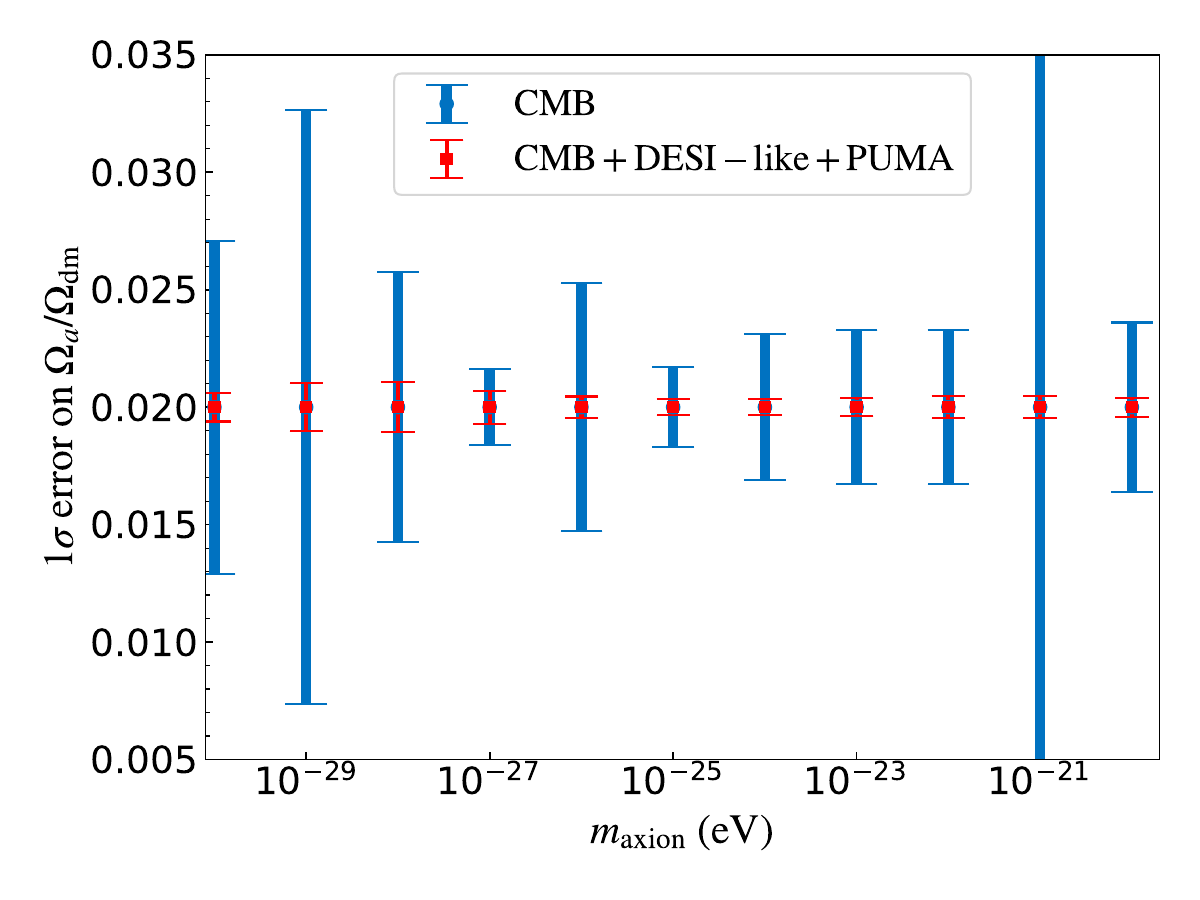}
    \caption{\textit{Projected 1$\sigma$ marginal errors on $\Omega_a/\Omega_{\rm dm}$ as a function of the ULA mass, assuming a fiducial value of $\Omega_a/\Omega_{\rm dm} = 0.02$. Each panel contrasts the constraints from CMB-S4-like missions alone against the joint analyses incorporating Ly$\alpha$, 21-cm auto-spectra, and their cross-spectra. The \textbf{left panel} utilises SKA1-MID for the 21-cm observations, while the \textbf{right panel} shows the results for PUMA. Both panels demonstrate an optimal detection window around $m_a \sim 10^{-25}$~eV, with the joint analyses yielding significantly tighter constraints than CMB-S4-like missions in isolation.}}
    \label{fig:omega_scf_cmb}
\end{figure}

Finally, we compare the projected 1$\sigma$ marginal errors on $\Omega_a/\Omega_{\rm dm}$ across the extended axion mass range $m_a \in [10^{-30}, 10^{-20}]$~eV. Fig.~\ref{fig:omega_scf_cmb} illustrates these estimated errors, contrasting the individual performance of CMB-S4-like setup against the joint CMB+LSS analyses for SKA1-MID (left panel) and PUMA (right panel). The joint analysis provides significantly reduced errors on the axion fraction, visibly surpassing the capabilities of standalone CMB-S4-like future experiments. Consistent with the trends observed in Fig.~\ref{fig:1sigma_omega_scf_wo_cmb}, the combined dataset exhibits an optimal sensitivity window for intermediate axion masses ($m_a \sim 10^{-25}$~eV), whereas the projected uncertainties at the extreme boundaries of the mass-window are comparatively large. Precisely, the  joint analysis of the Ly$\alpha$--21-cm cross-spectrum alongside CMB data represents a strong potential for probing the fraction of ULA dark matter. 

\begin{figure}[!ht]
    \centering
    \includegraphics[width=0.45\linewidth]{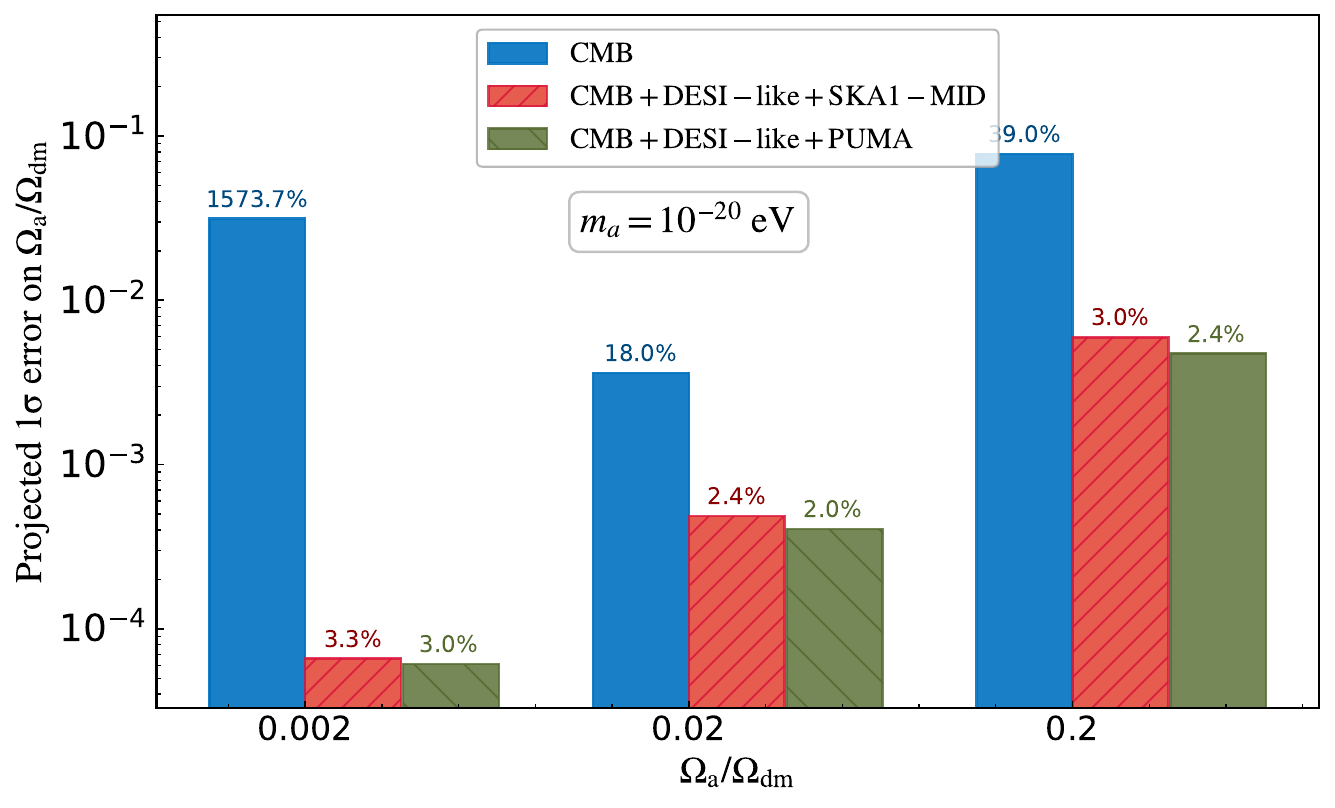}
    \hfill
    \includegraphics[width=0.45\linewidth]{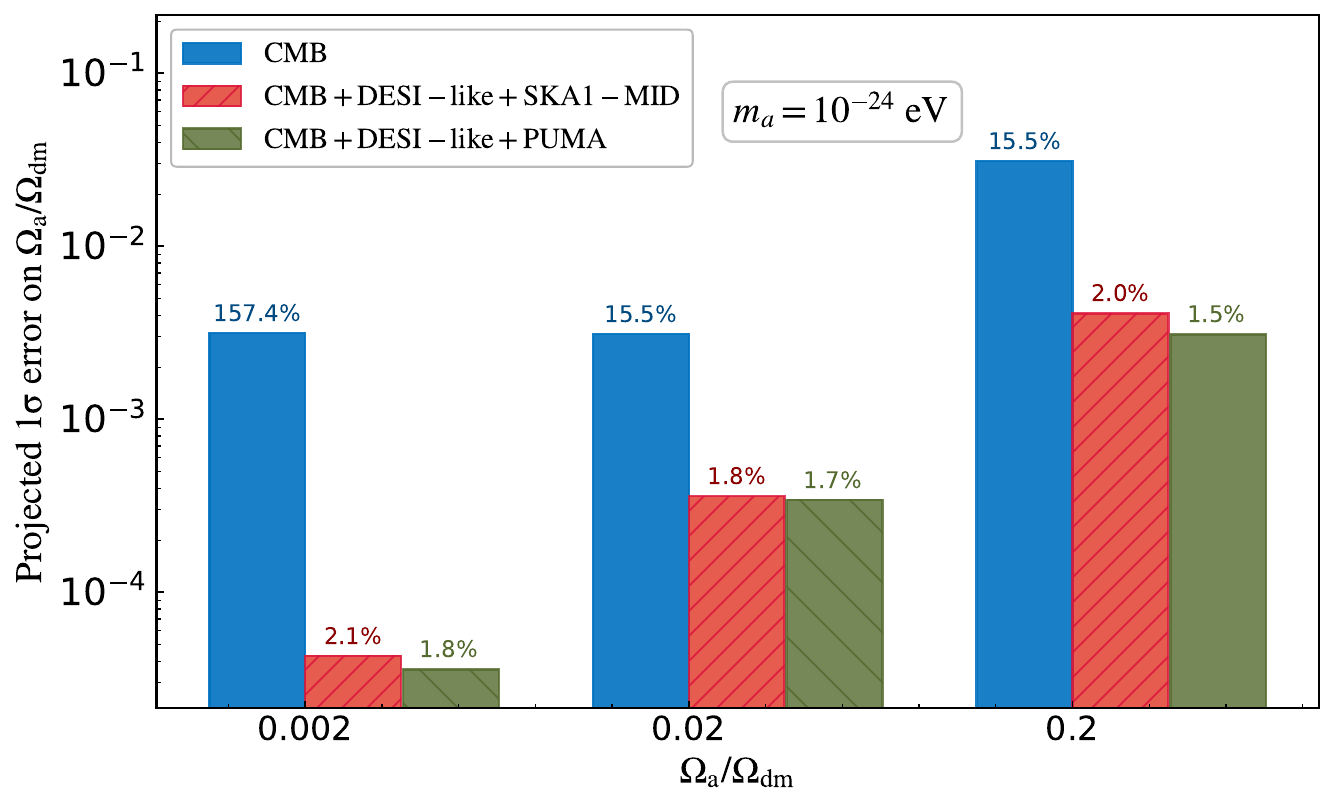}
    \includegraphics[width=0.45\linewidth]{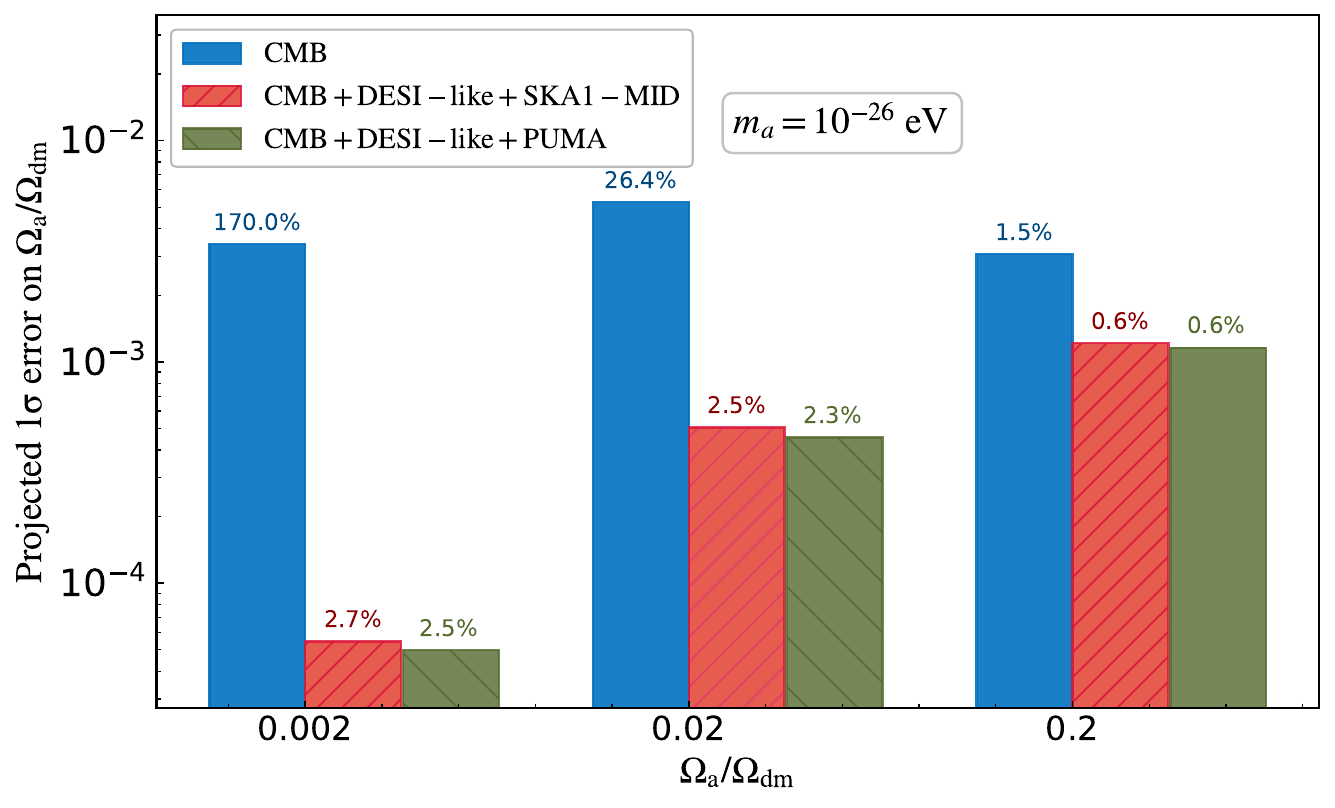}
    \hfill
    \includegraphics[width=0.45\linewidth]{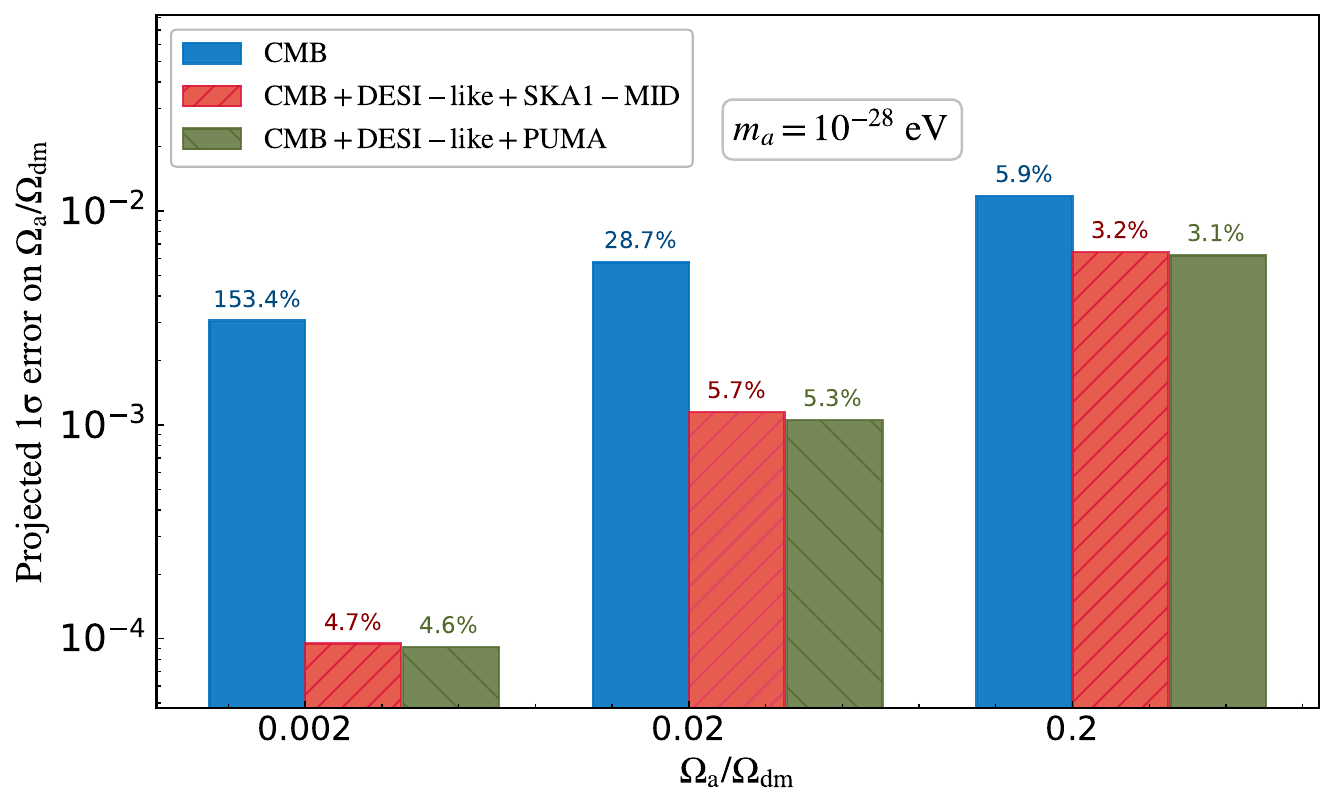}
    \caption{\textit{Projected $1\sigma$ uncertainties on the fractional ULA abundance ($\Omega_a/\Omega_{\rm dm}$) as a function of its several fiducial values, evaluated across four representative axion masses. The bar charts compare the constraining power of CMB-S4-like missions alone against the joint analyses (CMB-S4-like+DESI-like+SKA1-MID and CMB-S4-like+DESI-like+PUMA). The values displayed at the top of each bar indicate the corresponding relative percentage uncertainties. Across all mass scales and fiducial choices, the inclusion of post-reionization cross-correlations significantly tightens the detection prospects for the ULA fraction.}}
    \label{fig:omega_scf_vs_f}
\end{figure}

Let us now proceed to examine how these estimated uncertainties depend on the fiducial choice of the ULA fraction. Fig.~\ref{fig:omega_scf_vs_f} illustrates this  by considering three baseline choices for the axion fraction: $\Omega_a/\Omega_{\rm dm} \in \{0.2,\, 0.02,\, 0.002\}$, keeping the six standard cosmological parameters fixed to their values as in Table~\ref{tab:fiducial_parameters}. For each case, we compare the estimated $1\sigma$ uncertainties across three experimental configurations: CMB-S4-like setup alone, CMB-S4-like+DESI-like+SKA1-MID and CMB-S4-like+DESI-like+PUMA. The relative percentage uncertainties are displayed at the top of each corresponding bar and the summary of this exercise can be found in Table~\ref{tab:fisher_cmb_cross}. Our results demonstrate that combining CMB-S4-like experimental setup with post-reionization missions, in particular, the cross-correlations, significantly enhances the precision in the ULA fraction. This significant reduction in uncertainty persists across all the chosen fiducial values of ULA fraction and axion masses, thereby underscoring the importance of the joint analysis of upcoming CMB data and post-reionization cross-spectra as a potentially important probe for ULA dark matter. 

\begin{table}[!ht]
\centering
\scalebox{0.9}{%
\begin{tabular}{c c c c c}
\hline
\hline
Fiducial on $\frac{\Omega_a}{\Omega_{\rm dm}}$ & $m_a$ (eV) & CMB-S4-like & DESI+SKA+CMB-S4 & DESI+PUMA+CMB-S4 \\
\hline
\hline
\multirow{4}{*}{0.2} & $10^{-20}$ & $0.2\pm 0.078$ & $0.2\pm 0.0059$ & $0.2\pm 0.0047$ \\
& $10^{-24}$ & $0.2\pm 0.031$ & $0.2\pm 0.0041$ & $0.2\pm 0.0031$ \\
& $10^{-26}$ & $0.2\pm 0.0031$ & $0.2\pm 0.0012$ & $0.2\pm 0.0011$ \\
& $10^{-28}$ & $0.2\pm 0.012$ & $0.2\pm 0.0064$ & $0.2\pm 0.0062$ \\
\hline
\multirow{4}{*}{0.02} & $10^{-20}$ & $0.02\pm 0.0036$ & $0.02\pm 0.00048$ & $0.02\pm 0.00040$ \\
& $10^{-24}$ & $0.02\pm 0.0031$ & $0.02\pm 0.00036$ & $0.02\pm 0.00034$ \\
& $10^{-26}$ & $0.02\pm 0.0053$ & $0.02\pm 0.00050$ & $0.02\pm 0.00045$ \\
& $10^{-28}$ & $0.02\pm 0.0057$ & $0.02\pm 0.00115$ & $0.02\pm 0.00105$ \\
\hline
\multirow{4}{*}{0.002} & $10^{-20}$ & $0.002\pm 0.0315$ & $0.002\pm 0.000066$ & $0.002\pm 0.000061$ \\
& $10^{-24}$ & $0.002\pm 0.0031$ & $0.002\pm 0.000043$ & $0.002\pm 0.000036$ \\
& $10^{-26}$ & $0.002\pm 0.0034$ & $0.002\pm 0.000055$ & $0.002\pm 0.000050$ \\
& $10^{-28}$ & $0.002\pm 0.0031$ & $0.002\pm 0.000095$ & $0.002\pm 0.000091$ \\
\hline
\hline
\end{tabular}%
}
\caption{\textit{Fisher 1$\sigma$ constraints on the ULA fraction on its three different fiducial values for CMB-S4-like missions alone and in combination with the LSS cross-spectra, evaluated across four different axion masses.}}
\label{tab:fisher_cmb_cross}
\end{table}

\section{Summary and future directions}
\label{sec:discussion}

We conducted a comprehensive investigation into the observational signatures of ULA dark matter, focusing on its characteristic quantum pressure which suppresses gravitational clustering below the Jeans scale. To map this scale-dependent suppression across cosmic history, the  potential of post-reionization LSS tracers, specifically, the 21-cm auto-spectrum, the Ly$\alpha$ auto-spectrum and their cross-correlation, in conjunction with next-generation CMB observations have been explored. 

From the  SNR analysis, the cross-correlation technique has been found to be the most promising one so far as the prospects of measurements of the relevant parameters are concerned. Assuming a fiducial axion mass of $m_a = 10^{-24}\text{ eV}$ at a redshift of $z=2.33$ and a line-of-sight angle of $\mu=0.79$, the isolated auto-spectra yield substantial theoretical sensitivities. Our results have showed a peak SNR of $\sim 141$ for the DESI-like Ly$\alpha$ survey, approximately $95$ for SKA1-MID, and approximately $340$ for the advanced PUMA, for $\Omega_a/\Omega_{\rm dm}=0.2$. This SNR can be further enhanced if the modes align more closely with the line-of-sight. 
The cross-spectrum with DESI+SKA1-MID exhibits a peak-SNR of $\sim 70$ near the scale of $k \sim 0.8~h/{\rm Mpc}$, while the peak-SNR increases to $\sim 125$ for DESI+PUMA. Extending this analysis to track the redshift evolution ($z \in [2, 3]$), our findings reveal that the SNR peaks around $z \sim 2.1$ for DESI-like surveys. Interestingly, while the SNR for SKA1-MID decreases monotonically with redshift, PUMA exhibits a distinct sensitivity peak (see Fig.~\ref{fig:snr_z_profile}). Ultimately, the cross-correlation profiles for both radio setups with the DESI-like survey, peaking near $z \sim 2.1$.

In addition, we performed Fisher matrix analysis to estimate the uncertainties in the fractional abundance of ULA over ULA masses. Across the broad mass range of $10^{-30}\text{ eV} \le m_a \le 10^{-20}\text{ eV}$, the estimated error on ULA fraction is smaller for intermediate masses around $m_a \sim 10^{-25}\text{ eV}$. In this regime, the characteristic axionic Jeans scale falls within the mildly non-linear, high-SNR windows ($0.1 \lesssim k \lesssim 2~h/{\rm Mpc}$) of the LSS surveys. The improved specifications of PUMA yield reduced uncertainties on $\Omega_a/\Omega_{\rm dm}$ across the ULA mass range, thereby, making it a promising interferometric setup to probe $\Omega_a/\Omega_{\rm dm}$. Our results highlight that although individual experimental setup may project smaller uncertainties compared to cross-spectrum, analysing the cross-spectrum still represents a more realistic approach as it mitigates the individual inherent uncertainties due to systematics.

A salient feature of our Fisher matrix analysis is the synergy between future LSS surveys and CMB observations. While LSS data can constrain the ULA parameters with high precision, CMB measurements simultaneously constrain both ULA and primordial cosmological parameters. Combining these probes significantly improves the overall sensitivity.
Quantitatively, for a fiducial ULA fraction of $\Omega_a/\Omega_{\rm dm} = 0.02$ and mass $m_a=10^{-24}\text{ eV}$, CMB-S4-like future missions alone provide projected 1$\sigma$ uncertainties of $\pm 0.0031$. However, a joint analysis using DESI-like + SKA1-MID decreases these uncertainties to $\pm 0.00036$, and substituting SKA1-MID with PUMA further reduces the projected error to $\pm 0.00034$. 
Furthermore, this reduction in uncertainty holds consistently across different fiducial choices for the ULA fraction (see Fig.~\ref{fig:omega_scf_vs_f}). Thus, the joint analysis of CMB data alongside post-reionization cross-spectra emerges as a potentially novel probe for ULA dark matter.

Having said that, we must point out to a possible direction to carry the analysis forward. Although the combined SNR + Fisher forecast analyses highlight the potential of this multi-tracer approach, achieving this precision in practice needs a careful handling of several astrophysical foregrounds. Our future plan includes a comprehensive foreground modeling to handle several contaminations coming from possible astrophysical sources and generate a more realistic SNR therefrom followed by estimation of uncertainties. Moreover, to improve upon the analytical approximations for the highly non-linear regime, we aim to utilize high-resolution N-body and hydrodynamical simulations. This will help in capturing the non-linear evolution of the ULA field in a much better way, resulting in possible improvements in estimation of precision using future post-reionization, Lyman-$\alpha$ and CMB surveys.

\acknowledgments
The authors thank Arko Bhaumik, Purba Mukherjee and Rahul Shah for fruitful discussions. The authors also acknowledge the computational facilities  provided by the SyMeC HPC cluster of Indian Statistical Institute (ISI), Kolkata. DP thanks ISI, Kolkata for financial support through Senior Research Fellowship.  ADB acknowledges financial support from DST, India, under Grant No. IFA20-PH250 (INSPIRE Faculty Award). SP2 thanks ANRF,  Govt. of India, for partial support through Project No. CRG/2023/003984.

\bibliographystyle{bibi}
\bibliography{mybib.bib}
\end{document}